# Eigenspectra optoacoustic tomography achieves quantitative blood oxygenation imaging deep in tissues

Stratis Tzoumas[1,3†], Antonio Nunes[1,†], Ivan Olefir[1], Stefan Stangl[2], Panagiotis Symvoulidis[1,3], Sarah Glasl[1,3], Christine Bayer[2], Gabriele Multhoff[2,4], Vasilis Ntziachristos[1,3*]

[1] Institute for Biological and Medical Imaging (IBMI), Helmholtz Zentrum München, Neuherberg, Germany
[2] Department of Radiation Oncology, Klinikum rechts der Isar, Technische Universität München, München, Germany
[3] Chair for Biological Imaging, Technische Universität München. München, Germany
[4] CCG – Innate immunity in Tumor Biology, Helmholtz Zentrum München, Neuherberg, Germany

**Light propagating in tissue attains a spectrum that varies with location due to wavelength-dependent fluence attenuation by tissue optical properties, an effect that causes *spectral corruption*. Predictions of the spectral variations of light fluence in tissue are challenging since the spatial distribution of optical properties in tissue cannot be resolved in high resolution or with high accuracy by current methods. Spectral corruption has fundamentally limited the quantification accuracy of optical and optoacoustic methods and impeded the long sought-after goal of imaging blood oxygen saturation ($sO_2$) deep in tissues; a critical but still unattainable target for the assessment of oxygenation in physiological processes and disease. We discover a new principle underlying light fluence in tissues, which describes the wavelength dependence of light fluence as an affine function of a few reference base spectra, independently of the specific distribution of tissue optical properties. This finding enables the introduction of a previously undocumented concept termed eigenspectra Multispectral Optoacoustic Tomography (eMSOT) that can effectively account for wavelength dependent light attenuation without explicit knowledge of the tissue optical properties. We validate eMSOT in more than 2000 simulations and with phantom and animal measurements. We find that eMSOT can quantitatively image tissue $sO_2$ reaching in many occasions a better than 10-fold improved accuracy over conventional spectral optoacoustic methods. Then, we show that eMSOT can spatially resolve $sO_2$ in muscle and tumor; revealing so far unattainable tissue physiology patterns. Last, we related eMSOT readings to cancer hypoxia and found congruence between eMSOT tumor $sO_2$ images and tissue perfusion and hypoxia maps obtained by correlative histological analysis.**

The assessment of tissue oxygenation is crucial for understanding tissue physiology and characterizing a multitude of conditions including cardiovascular disease, diabetes, cancer hypoxia[1] or metabolism. Today, tissue oxygenation ($pO_2$) and hypoxia measurements remain challenging and often rely on invasive methods that may change the tissue physiology, such as single point needle polarography or immunohistochemistry[2]. Non-invasive imaging methods have been also considered, underscoring the importance of assessing $pO_2$, but come with limitations. Positron emission tomography (PET) or single-photon emission computed tomography (SPECT) assess cell hypoxia by administration of radioactive tracers[2], but are often not well suited for quantifying tissue oxygenation, suffer from low spatial resolution and are unable to provide longitudinal or dynamic imaging capabilities. Electron paramagnetic resonance imaging[3] can measure tissue $pO_2$ but is not widely used due to limitations in spatial and temporal resolution. Imaging methods using tracers may be further limited by restricted tracer bio-distribution, in particular to hypoxic areas. Tracer-free modalities have also been researched, in particular BOLD MRI[4], which however primarily assesses only deoxygenated hemoglobin and therefore presents challenges in quantifying oxygenation and blood volume[5].

Measurement of blood oxygenation levels ($sO_2$) is a vital tissue physiology measurement and can provide an alternative way to infer tissue oxygenation and hypoxia. Arterial $sO_2$ is widely assessed by the pulse oximeter, based on empirical calibrations, but this technology cannot be applied to measurements other than arterial blood. Optical microscopy methods like phosphorescence quenching microscopy[6] or optoacoustic (photoacoustic) microscopy[7] can visualize oxygenation in blood vessels and capillaries but are restricted to superficial (<1mm depth) measurements. Diffuse optical methods received significant attention in the last two decades for sensing and imaging oxy- and deoxygenated hemoglobin deeper in tissue but did not yield sufficient accuracy because of the low resolution achieved due to photon scattering[8].

Multispectral optoacoustic tomography (MSOT) detects the spectra of oxygenated and deoxygenated hemoglobin in high resolution deep within tissues, since signal detection and image reconstruction are not significantly affected by photon scattering [9,10]. Despite the principal MSOT suitability for non-invasive imaging of blood oxygenation, accuracy remains limited by the dependence of light fluence on depth and light color. Unless explicitly accounted for, the wavelength dependent light fluence attenuation with depth alters the spectral features detected and results in inaccurate estimates of blood $sO_2$ [11,12]. Despite at least two decades of research in optical imaging, the problem of light fluence correction has not been conclusively solved [9]. To date this problem has been primarily studied from an optical property quantification point of view [13,12]. However, it is not possible today to accurately image tissue optical properties *in-vivo*, in high-resolution, and compute light fluence[12]. Therefore, quantitative $sO_2$ measurement deep in tissue *in-vivo* remains an unmet challenge. Conventional spectral optoacoustic methods[14,15] typically ignore the effects of light fluence and employ linear spectral fitting with the spectra of oxy- and deoxy-hemoglobin for estimating sO$_2$ (linear unmixing), a common simplification that can introduce substantial errors in deep tissue.

[†] Equal author contribution
[*] Correspondence to V.N. (v.ntziachristos@tum.de)

In this work we discovered a new principle of light fluence in tissue and exploit it to solve this fundamental quantification challenge of optical methods. In particular, we found that the spectral patterns of light fluence expected within the tissue can be modeled as an affine function of a few reference base spectra, independently of the specific distribution of tissue optical properties or the depth of the observation. We show how this principle can be employed to solve the spectral corruption problem without knowledge of the tissue optical properties, and significantly increase the accuracy of spectral optoacoustic methods. The proposed method, termed eigenspectra-MSOT (eMSOT), provides for the first time quantitative estimation of blood sO$_2$ in deep tissue. We demonstrate the superior performance of the method with more than 2000 simulations, phantom measurements and *in-vivo* controlled experiments. Then, using eMSOT, we image for the first time oxygen gradients in skeletal muscles *in-vivo*, previously only accessible through invasive methods. Furthermore, we show the application of eMSOT in quantifying blood oxygenation gradients in tumors during tumor growth or O$_2$ challenge and relate label-free non-invasive eMSOT readings to tumor hypoxia; demonstrating the ability to measure quantitatively the perfusion hypoxia level in tumors, as confirmed with invasive histological gold standards.

## RESULTS

A new concept of treating light fluence in diffusive media/tissues is introduced, based on the observation that the light fluence spectrum at different locations in tissue depends on a cumulative light absorption operation by tissue chromophores, such as hemoglobin. We therefore hypothesized that there exists a small number of base spectra that can be combined to predict any fluence spectrum present in tissue; therefore avoiding the unattainable task of knowing the distribution of tissue optical properties at high resolution. To prove this hypothesis, we first applied Principal Component Analysis (PCA) on 1470 light fluence spectral patterns, which were computed by simulating light propagation in tissues at 21 different (uniform) oxygenation states of hemoglobin and 70 different discrete depths (**Methods**). PCA analysis yielded four significant base spectra, i.e. a mean light fluence spectrum (**Figure 1a**) and three fluence *Eigenspectra* (**Figure 1b-d**).

We then postulated that light fluence spectra in unknown and non-uniform tissues can be modeled as a superposition of the mean fluence spectrum ($\Phi_M$) and the three *Eigenspectra* ($\Phi_i(\lambda)$, i=1..3) multiplied by appropriate scalars m$_1$, m$_2$ and m$_3$, termed *Eigenfluence* parameters. To validate this hypothesis we computed the light fluence in >500 simulated tissue structures of different and non-uniform optical properties and hemoglobin oxygenation values (**Supplementary Note 1**). For each pixel, we fitted the simulated light fluence spectrum to the *Eigenspectra* model and derived the *Eigenfluence* parameters (m$_1$, m$_2$, m$_3$) and a fitting residual value. The residual value represents the error of the *Eigenspectra* model in matching the simulated data and typically assumed values below 1% (see **Supplementary Note 1**) indicating that three *Eigenspectra* can accurately model all simulated fluence spectra generated in tissues of arbitrary structure. We further observed that the values of m$_2$ vary primarily with tissue depth while the values of m$_1$ also depend on the average levels of background tissue oxygenation (see **Figure 1f-h**). Intuitively this indicates that the second *Eigenspectrum* $\Phi_2(\lambda)$ is mainly associated with the modifications of light fluence spectrum due to depth and the average optical properties of the surrounding tissue, while the first *Eigenspectrum* $\Phi_1(\lambda)$ is also associated with the "spectral shape" of light fluence that relates to the average oxygenation of the surrounding tissue.

Following these observations, we propose eigenspectra MSOT (eMSOT), based on three eigenspectra $\Phi_1(\lambda)$, $\Phi_2(\lambda)$, $\Phi_3(\lambda)$, as a method that formulates the blood sO$_2$ estimation problem as a non-linear spectral unmixing problem, i.e.

$$P(\mathbf{r},\lambda) = \Phi'(\mathbf{r},\lambda) \cdot (c'_{HbO2}(\mathbf{r}) \cdot \varepsilon_{HbO2}(\lambda) + c'_{Hb}(\mathbf{r}) \cdot \varepsilon_{Hb}(\lambda)), \qquad (1)$$

where P($\mathbf{r}$,$\lambda$) is the multispectral optoacoustic image intensity obtained at a position $\mathbf{r}$ and wavelength $\lambda$, $\varepsilon_{HbO2}(\lambda)$ and $\varepsilon_{Hb}(\lambda)$ are the wavelength dependent molar extinction coefficients of oxygenated and deoxygenated hemoglobin, c′$_{HbO2}$($\mathbf{r}$) and c′$_{Hb}$($\mathbf{r}$) are the relative concentrations of oxygenated and deoxygenated hemoglobin (proportional to the actual ones with regard to a common scaling factor, see **Methods**), and $\Phi'(\mathbf{r},\lambda) = \Phi_M(\lambda) + m_1(\mathbf{r})\Phi_1(\lambda) + m_2(\mathbf{r})\Phi_2(\lambda) + m_3(\mathbf{r})\Phi_3(\lambda)$. Eq. (1) defines a non-linear inversion problem, requiring measurements at 5 wavelengths or more for recovering the 5 unknowns, i.e. c'$_{HbO2}$($\mathbf{r}$), c'$_{Hb}$($\mathbf{r}$), m$_1$($\mathbf{r}$), m$_2$($\mathbf{r}$), m$_3$($\mathbf{r}$) and is solved as a constrained optimization problem (**Methods, Supplementary Note 2**). For computational efficiency, we observe that the light fluence varies smoothly in tissue and only compute the *Eigenfluence* parameters on a coarse grid subsampling the region of interest (**Figure 1i**). Then, cubic interpolation is employed to compute the *Eigenfluence* parameters in each pixel within the convex hull of the grid (**Figure 1j**) and calculate a fluence spectrum $\Phi'(\mathbf{r},\lambda)$ for each pixel. Eq. (1) is then solved for c′$_{HbO2}$($\mathbf{r}$) and c′$_{Hb}$($\mathbf{r}$) and sO$_2$ is computed (see **Methods**).

Using simulated data obtained from a light propagation model (finite element solution of the diffusion approximation) applied on >2000 randomly created maps of different optical properties, simulating different tissue physiological states, we found a substantially improved sO$_2$ estimation accuracy of eMSOT over linear unmixing (**Figure 1m & Supplementary Note 3**). Especially in the case of tissue depths of >5mm eMSOT typically offered a 3-8 fold sO$_2$ estimation accuracy improvement over conventional linear unmixing (**Supplementary Figure 4n**). **Figure 1k** depicts a representative example of a simulated blood sO$_2$ map and visually showcases the differences between the eMSOT sO$_2$ image (middle), the sO$_2$ image obtained using linear unmixing (left) and the original sO$_2$ simulated image (right). eMSOT offered significantly lower sO$_2$ estimation error with depth, compared to the linear fitting method (**Figure 1l**).

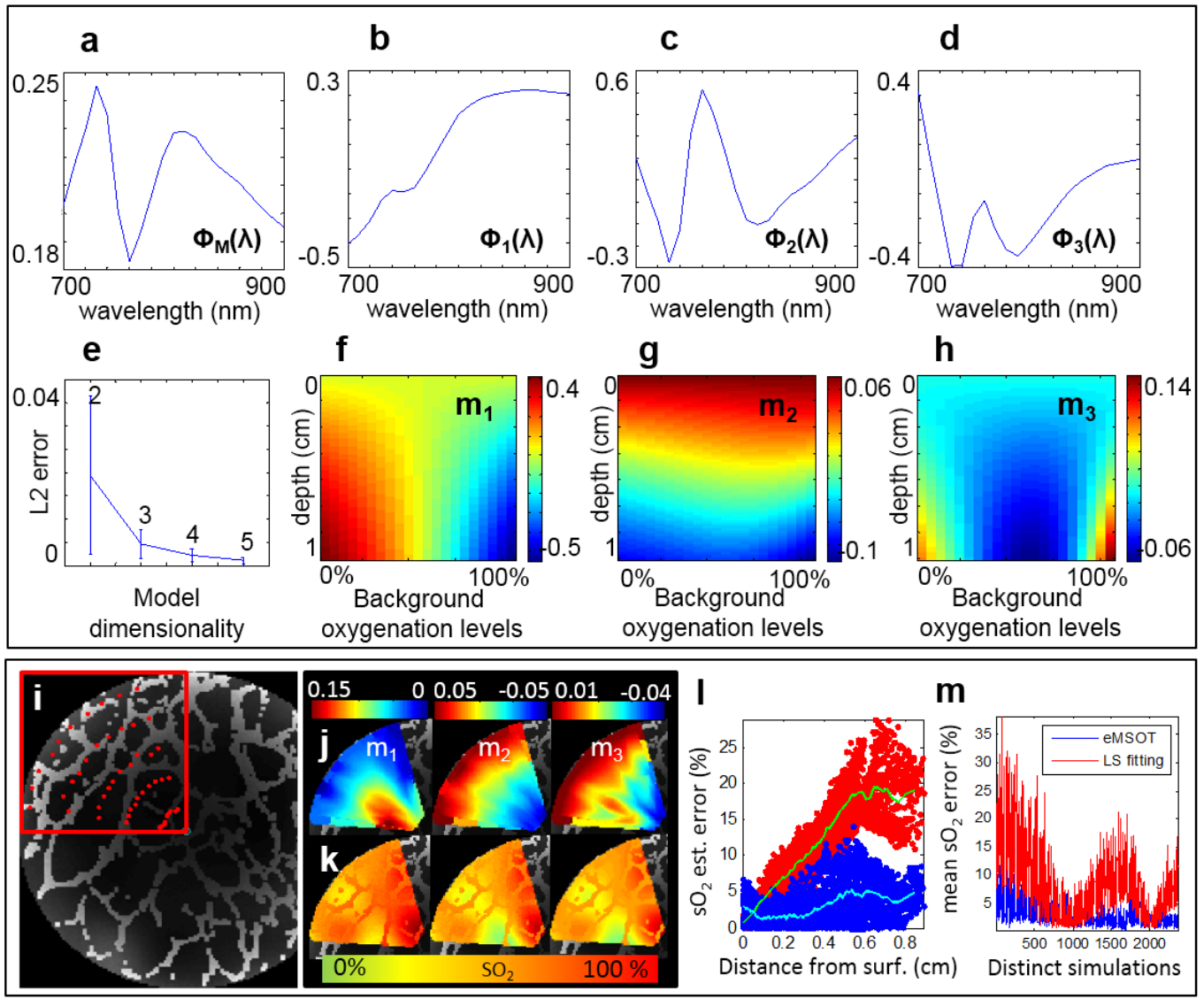

**Figure 1. eMSOT concept and application. (a-d)** The *Eigenspectra* model composed of a mean fluence spectrum $\Phi_M(\lambda)$ **(a)** and the three fluence *Eigenspectra* $\Phi_1(\lambda)$, $\Phi_2(\lambda)$ and $\Phi_3(\lambda)$, **(b), (c), (d),** respectively, as derived by applying PCA on a selected training data-set of light fluence spectra. **(e)** L2 norm error of the *Eigenspectra* model on the training dataset for different model dimensionalities. **(f-h)** Values of the parameters $m_1$, $m_2$ and $m_3$ as a function of tissue depth (y axis) and tissue oxygenation (x axis). The values have been obtained after fitting the light fluence spectra of the training data-set (see Methods) to the *Eigenspectra* model. **(i)** Application of a circular grid (red points) for eMSOT inversion on an area of a simulated MSOT image. **(j)** After eMSOT inversion the model parameters $m_1$, $m_2$ and $m_3$ are estimated for all grid points and maps of $m_1$, $m_2$ and $m_3$ are produced for the convex hull of the grid by means of cubic interpolation. These maps are used to spectrally correct the original MSOT image. **(k)** Blood $sO_2$ estimation using linear unmixing (left), eMSOT (middle) and Gold standard $sO_2$ (right) of the selected region. **(l)** $sO_2$ estimation error of the analyzed area sorted per depth in the case of linear unmixing (red points) and eMSOT (blue points). **(m)** Mean $sO_2$ error of linear unmixing (red) and eMSOT (blue) corresponding to >2000 simulations of random structures and optical properties (see Supplementary Note 3).

For experimentally assessing the accuracy of eMSOT, we performed a series of blood phantom experiments that suggest an up to 10-fold more reliable $sO_2$ estimation derived by eMSOT, as compared to conventional linear unmixing (**Supplementary Note 4**). In addition, controlled mouse measurements (n=4) were performed *in-vivo*, under gas anesthesia, by rectally inserting capillary tubes containing blood at 100% and 0% $sO_2$ levels (**Methods**). The mice were imaged in the lower abdominal area under 100%$O_2$ and 20% $O_2$ breathing conditions (**Figure 2a**). Figure 2a showcases the eMSOT grid applied on the images processed (left column), the $sO_2$ maps obtained with linear unmixing (middle column) and with eMSOT (right column). The spectral fitting of linear unmixing (left) and eMSOT (right) corresponding to a pixel in the area of the capillary tube (yellow arrows in **a**) are presented in **Figure 2b** along with the estimated $sO_2$ values. In the controlled *in-vivo* experiments, the mean linear unmixing error ranged from 16 to 35% while eMSOT offered a mean $sO_2$ error ranging from 1 - 4% indicating an order of magnitude improved accuracy (**Figure 2c**).

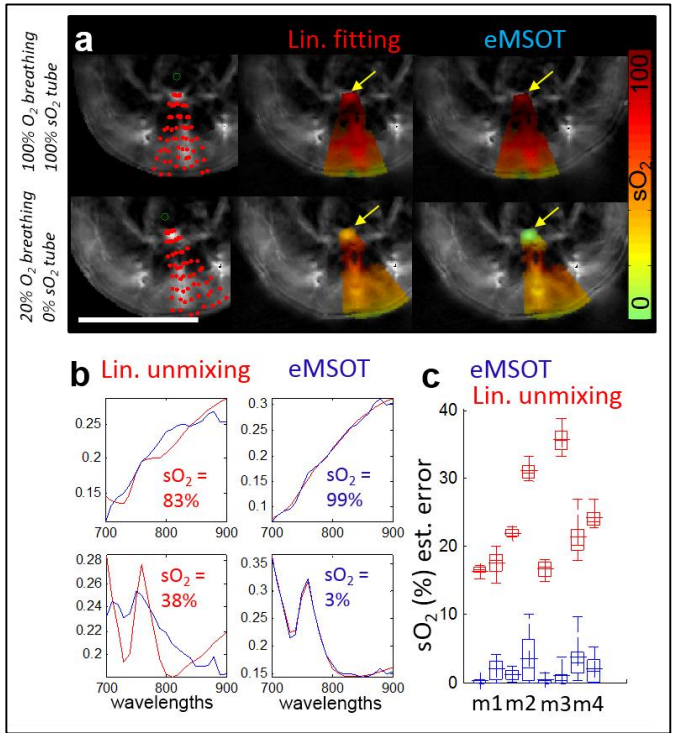

**Figure 2. Comparison of eMSOT sO$_2$ estimation accuracy over conventional spectral optoacoustic method.** (**a**) eMSOT application in the case of *in-vivo* controlled experiments under 100% O$_2$ (**a** upper row) and 20% O$_2$ (**a** lower row) breathing. Capillary tubes containing blood of 100% sO$_2$ (upper row) and 0% sO$_2$ (lower row) were inserted within tissue (arrows). Scale bar 1cm. (**b**) Spectral fitting and sO$_2$ estimation in the insertion area (yellow arrows in **a**) using linear unmixing (left column) and eMSOT (right column). The blue curves correspond to P(**r**,λ) (left column) and P$^{eMSOT}$(**r**,λ) (right column) while the red curves correspond to $c_{HbO2}^{lu}(\mathbf{r})\varepsilon_{HbO2}(\lambda)+c_{Hb}^{lu}(\mathbf{r})\varepsilon_{Hb}(\lambda)$ (left column; the term lu refers to linear unmixing) and $c'_{HbO2}^{eMSOT}(\mathbf{r})\varepsilon_{HbO2}(\lambda)+c'_{Hb}^{eMSOT}(\mathbf{r})\varepsilon_{Hb}(\lambda)$ (right column). (**c**) sO$_2$ estimation error using eMSOT (blue) and linear unmixing (red) in all four animal experiment repetitions.

Blood oxygenation and oxygen exchange in the microcirculation have been traditionally studied through invasive, single-point polarography or microscopy measurements in vessels and capillaries of the skeletal muscle[16]. Research for macroscopic methods that could non-invasively resolve muscle oxygenation was broadly pursued in the past two decades by considering Near-Infrared Spectroscopy (NIRS) and Diffuse Optical Tomography (DOT) [17,18], which, however did not produce solutions yielding high fidelity or resolution. In a next set of experiments we, therefore, studied whether eMSOT could non-invasively quantify the oxygenation gradient in the skeletal muscle and we compared this performance to conventional spectral optoacoustic methods. The hindlimb muscle of 6 nude mice was imaged *in-vivo* under 100% O$_2$, 20% O$_2$ challenge; three of the mice were then sacrificed with an overdose of CO$_2$, the latter binding to hemoglobin and deoxygenating blood.

eMSOT resolved oxygenation gradients in the muscle, as a function of breathing conditions *in-vivo* (**Figure 3 b-c**) and *post-mortem* after CO$_2$ breathing (**Figure 3d**). The *post-mortem* deoxygenated muscle served herein as a control experiment and was also analyzed with linear unmixing for comparison (**Figure 3e**). In the *post-mortem* case, linear unmixing overestimated the sO$_2$ as a function of tissue depth (**Figure 3e**) and yielded large errors in matching the tissue spectra (**Figure 3f** – upper row). Conversely, eMSOT offered sO$_2$ measurements in agreement with the expected physiological states (**Figure 3b-d**) and consistently low fitting residuals (**Figure 3f** – lower row, **Supplementary Figure 6**). **Figure 3d-e** and **Figure 3f** demonstrate the prominent effects of spectral corruption with depth that impair the accuracy of conventional spectral optoacoustic methods but are tackled by eMSOT. The estimated blood sO$_2$ values corresponding to a deep tissue area (yellow rectangle in **Figure 3b**) are tabulated in **Figure 3g** for eMSOT and linear unmixing and depict that the latter demonstrated unrealistically small sO$_2$ changes between the normoxic *in-vivo* and anoxic *post-mortem* (after CO$_2$ breathing) states.

In addition to physiological tissue features, MSOT also reveals tissue morphology. MSOT images at a single wavelength (900 nm) captured prominent vascular structures likely corresponding to femoral vessels or their branches (**Figure 3h**) with implicitly co-registered eMSOT blood-oxygenation images. This hybrid mode enables the study of physiology at specific tissue areas. We selected to study blood oxygenation measurements at a region of interest around large vessels (ROI-1; **Figure 3h**) and a region of interest within the muscle presenting no prominent vascular structures (ROI-2; **Figure 3h**) for the 100% O$_2$, 20% O$_2$ and CO$_2$ breathing conditions. Average tissue sO$_2$ was typically measured at 60%-70% saturation under medical air breathing and at 70%-80% saturation under 100%O$_2$ breathing near large vessels (**Figure 3j**). Average tissue blood oxygenation away from large vessels (ROI-2) was estimated consistently lower, at 35 - 50% saturation under normal breathing conditions and 45-60% saturation under 100%O$_2$ breathing (**Figure 3k**). These are first observations of quantitative high-resolution blood-oxygenation spatial gradients imaged non-invasively in tissue.

The low blood saturation values in tissue (35 -50%) cannot be explained by considering arterial and venous blood saturation. However, previous studies based on direct microscopy measurements in vessels and capillaries through polarography, hemoglobin spectrophotometry and phosphorescence quenching microscopy have revealed similar oxygenation gradient in the skeletal muscle [16] with hemoglobin saturation in the femoral artery found to range between 87-99% sO$_2$[16,19], while rapidly dropping down to 50-60% sO$_2$ in smaller arterioles [19,20]. The average oxygen saturation in venules and veins has been found to range between 45%-60% sO$_2$ under normal breathing conditions, reaching up to 70% at 100% O$_2$ breathing [20,21]. Average capillary blood oxygenation has been estimated at 40% sO$_2$ with a large standard deviation [21], often reported lower, at an average, than venular oxygenation[16]. Therefore, the eMSOT values measured at ROI-1 possibly relate to a weighted average of arterial/arteriolar and venous/venular sO$_2$ in skeletal muscle, while the values measured at ROI-2, which anatomically presents no prominent vasculature, relate more to capillary sO$_2$ measurements.

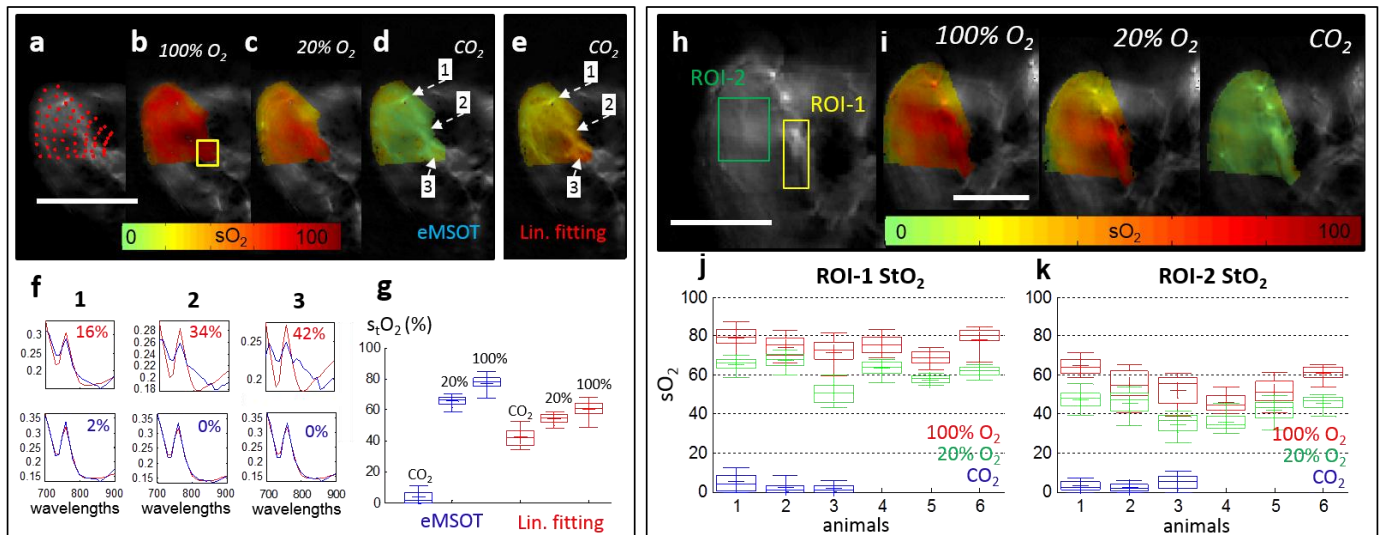

**Figure 3. eMSOT measurements of tissue blood oxygenation in the muscle.** (**a-d**) eMSOT grid applied on the area of the hindlimb muscle (**a**) and eMSOT tissue blood $sO_2$ estimation in the case of 100% $O_2$ breathing (**b**), 20% $O_2$ breathing (**c**) and *post-mortem* after $CO_2$ breathing (**d**). (**e**) $sO_2$ estimation using linear unmixing in the *post-mortem* case after $CO_2$ breathing. Scale bar 1cm. (**f**) Spectral fitting and $sO_2$ values of linear unmixing (upper row) and eMSOT (lower row) for the three points indicated in (**d**) and (**e**). The blue curves correspond to $P(\mathbf{r},\lambda)$ (upper row) and $P^{eMSOT}(\mathbf{r},\lambda)$ (lower row) while the red curves correspond to $c_{HbO2}^{lu}(\mathbf{r})\varepsilon_{HbO2}(\lambda)+c_{Hb}^{lu}(\mathbf{r})\varepsilon_{Hb}(\lambda)$ (upper row) and $c'_{HbO2}{}^{eMSOT}(\mathbf{r})\varepsilon_{HbO2}(\lambda)+c'_{Hb}{}^{eMSOT}(\mathbf{r})\varepsilon_{Hb}(\lambda)$ (lower row). (**g**) Estimated blood $sO_2$ of a deep tissue area (yellow box in **b**) using eMSOT (blue) and linear unmixing (red). (**h**) Anatomical MSOT image of the hindlimb area at an excitation wavelength of 900 nm. Two regions were selected for presenting the $sO_2$ values, one close to prominent vasculature (ROI-1) and one corresponding to soft tissue (ROI-2). Scale bar 0.5cm. (**i**) eMSOT $sO_2$ estimation *in-vivo* under 100% (left) and 20% $O_2$ breathing (middle) and *post-mortem* after $CO_2$ breathing (right). Scale bar 0.5cm. (**j, k**) Estimated tissue $sO_2$ of ROI-1 (**j**) and ROI-2 (**k**) under 100% (red) and 20% $O_2$ breathing (green) and *post-mortem* after $CO_2$ breathing (blue). Measurements correspond to 6 different animals.

The improved accuracy observed in eMSOT over previous approaches and general agreement with invasive tissue measurements prompted the further study of perfusion hypoxia emerging from the incomplete delivery of oxygenated hemoglobin in tissue areas. We hypothesized that measurements of blood saturation could be employed as a measure of tissue hypoxia, assuming natural hemoglobin presence in hypoxic areas. To examine this hypothesis we applied eMSOT to measure blood oxygenation in 4T1 solid tumors orthotopically implanted in the mammary pad of 8 mice (**Methods, Supplementary Note 6**). MSOT revealed the tumor anatomy and heterogeneity, which was found consistent to anatomical features identified through cryoslice color photography and H&E staining (**Supplementary Note 6**). Furthermore, imaging tumors at different time-points revealed the progression of hypoxia during tumor growth (**Figure 4a-b**). The spread of hypoxia, i.e. the presence of hypoxic area under a threshold (varied from 50% to 25% $sO_2$) over the total tumor area, also increased during tumor progression (**Figure 4c**). Following the *in-vivo* measurements we harvested the tumor tissue and related the non-invasive eMSOT findings to the histological assessment of tumor hypoxia (see **Supplementary Note 6**). Tumor tissue was stained by Hoechst 33342 [22] (indicating perfusion) and Pimonidazole [23] (indicating cell hypoxia). The results indicated close correspondence between the hypoxic areas detected by eMSOT using hemoglobin as a hypoxia sensor (**Figure 4b**) and the histology slices (**Figure 4d**). We found that eMSOT could not only quantitatively distinguish between high and low hypoxia levels in the tumors, but the spatial $sO_2$ maps further presented congruence with the spatial appearance of hypoxia spread and reduced perfusion seen in the histology slices (**Figure 4e-g**). A quantitative congruence analysis is shown in **Supplementary Note 6**. Finally, clear differences were also observed between the hypoxic tumor and healthy tissue response to an $O_2$ breathing challenge (**Figure 4h; Supplementary Figure 8**), with areas in the core of the tumor presenting a limited response to such external stimuli, likely due to the presence of non-functional vasculature.

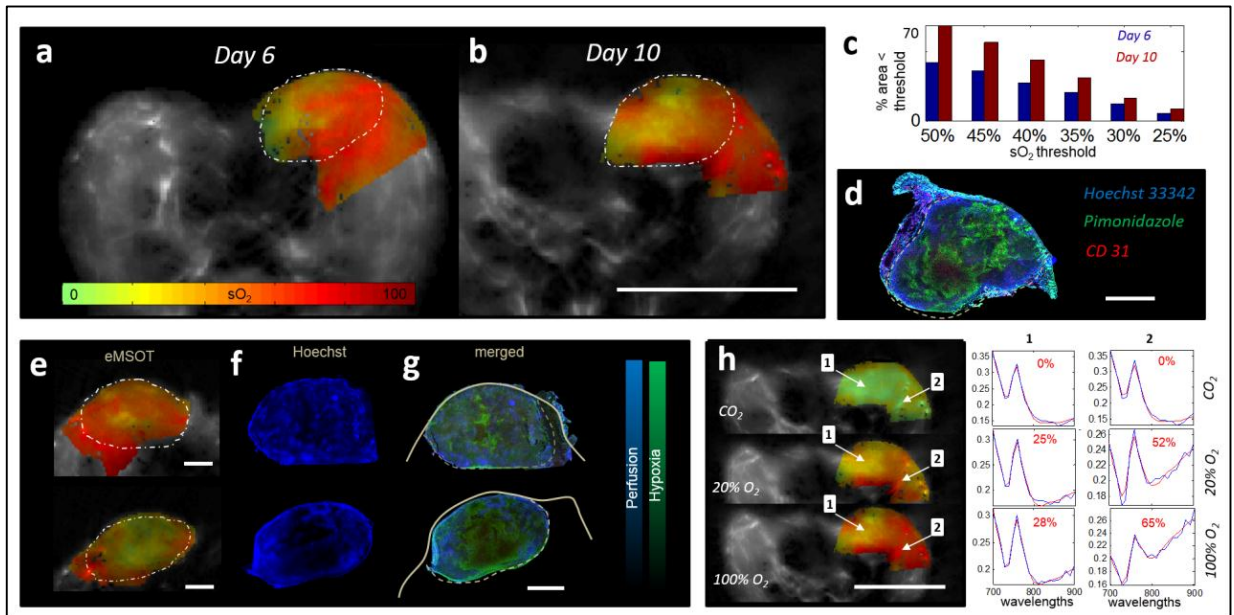

**Figure 4. eMSOT measurements of tissue blood oxygenation in tumor.** (**a-b**) $sO_2$ maps of a 4T1 tumor implanted in the mammary pad at day 6 (**a**) and day 10 (**b**) after cell inoculation. Dashed lines present a segmentation of the tumor area. Scale bar 1cm. (**c**) Bar-plot presenting the percentage of the total tumor area containing $sO_2$ values lower than a specific $sO_2$ threshold (x axis). Blue bars correspond to the tumor imaged at day 6 and red bars correspond to the tumor imaged at day 10, presented in (**a, b**). (**d**) Merged Hoechst 33342 and Pimonidazole staining of the tumor presented in (**b**). Scale bar, 2mm. (**e-g**) Examples of a highly perfused (upper row) and low perfused (lower row) tumor analysed with eMSOT for $sO_2$ estimation (**e**), Hoechst 33342 staining (**f**), and merged with Pimonidazole staining (**g**). Hoechst staining presented lower intensity at tumors and tumor areas presenting low $sO_2$ values, as measured by eMSOT. Scale bar, 2mm. (**h**) $sO_2$ map of a tumor under an $O_2$–$CO_2$ challenge. The computed $sO_2$ values and the eMSOT spectral fit of points 1 and 2 (arrows) are presented in (**h** right) for the three breathing conditions. Scale bar 1cm.

## DISCUSSION

Spectral corruption has so far limited the potential of optical and optoacoustic methods to offer accurate, quantitative assessment of blood oxygen saturation deep inside tissues. Conventional computational methods in optical imaging propose to invert a light transport operator to recover tissue optical properties (absorption and scattering) [12]; then use these properties for calculating tissue physiological parameters. However, the complexity and ill-posed nature of the inversion problem has not allowed so far accurate, high-resolution $sO_2$ imaging. We discovered a new principle that describes the spectral features of light fluence as a combination of spectral base functions. Using this principle, we formulated the $sO_2$ quantification problem as a non-linear spectral unmixing problem that does not require knowledge of tissue optical properties. Effectively, eMSOT converts $sO_2$ imaging from a problem that is spatially dependent on light propagation and optical properties, as common in traditional optical methods, to a problem solved in the spectral domain. Therefore, $sO_2$ can be directly quantified without estimating tissue optical properties.

eMSOT requires theoretically at least 5 excitation wavelengths for resolving spectral domain parameters and the relative oxygenated and deoxygenated hemoglobin concentrations. We hereby utilized 21 wavelengths for ensuring high accuracy. The recent evolution of video-rate MSOT imaging systems, based on fast tuning optical parametric oscillator lasers[24] allows the practical implementation of the method. Modern MSOT systems offer 5 wavelength scans at 20Hz or better. Therefore eMSOT is a technology that optimally interfaces to a new generation of fast and handheld spectral optoacoustic systems [25].

The method developed demonstrated quantitative, non-invasive blood oxygenation images in phantoms and tissues *in-vivo* (muscle and tumor) in high-resolution, showing good correlation with the expected physiological state or the histologically observed spatial distribution of perfusion and hypoxia. eMSOT measures blood oxygenation. We hypothesized that a correlation exists to tissue oxygenation measurements by assuming a wide presence of hemoglobin in tissues. We demonstrated congruence (**Supplementary Note 6**) between traditional invasive histological assays resolving tissue hypoxia and eMSOT analysis. Importantly, not only average values are resolved, but there is a close spatial correspondence between hypoxia patterns resolved by eMSOT non-invasively and histological analysis (**Figure 4, Supplementary Figure 7**).

High-resolution non-invasive imaging of blood oxygenation across entire tissues and tumors offers novel abilities in studying physiological and pathological conditions. This goal has been pursued for decades with near-infrared methods, but the strong effects of photon scattering and photon diffusion on the signals detected limited the imaging resolution and impeded accurate quantification[8]. Optoacoustic imaging improves the resolution achieved, over diffuse optical imaging methods but its $sO_2$ estimation accuracy has been limited so far by depth-dependent fluence attenuation and spectral corruption effects. We showed that conventional spectral optoacoustic methods employing linear unmixing can significantly misestimate blood saturation values in several controlled measurements, including simulations and animal measurements. eMSOT was tested on a vast data-set consisting of >2000 tissue simulations and was consistently found to provide from a comparable to substantially better $sO_2$ estimation accuracy over linear unmixing. (**Supplemental Note 3**). The large number of simulations was necessary to validate eMSOT, which presents a non-convex optimization problem. eMSOT was further tested on tissue mimicking blood phantoms (**Supplemental Note 4**) and controlled *in-vivo* experiments (**Figure 2, Supplemental Note 5**). In all cases tested, eMSOT offered from comparable to significantly more accurate performance over conventional spectral optoacoustic methods.

A particular challenge in this study was the confirmation of the eMSOT values obtained *in-vivo*. Polarography measurements are invasive, disrupt the local microenvironment and do not allow to recover spatial information. Nuclear methods using tracers are not well suited for longitudinal studies and utilize tracers which need to distribute in hypoxia areas i.e. areas with problematic supply. Therefore the results may not directly compare to eMSOT, even though such study is planned as a next step. BOLD MRI only indirectly resolves the effects of deoxygenated hemoglobin but cannot observe oxygenated hemoglobin. For this reason, we selected to utilize traditional histology methods, using cryoslicing, which allows to maintain spatial orientation so that eMSOT and histological results could be compared not only in terms of quantity but also in regard to the spatial appearance.

eMSOT offers a novel solution to a fundamental challenge in optical and optoacoustic imaging. In the absence of other reliable methods that can image blood oxygenation, it may be that eMSOT becomes the gold standard method in blood and tissue oxygenation studies. Its congruence with tissue hypoxia may also allow a broad application in tissue and cancer hypoxia studies. Nevertheless eMSOT performs optimally when applied on well-reconstructed parts of optoacoustic images (**Supplementary Note 5**). For this reason, it was selectively applied herein to the part of the image that is within the optimal sensitivity field of the detector employed. An eMSOT advantage is that it is insensitive to scaling factors such as the Grüneisen coefficient or the spatial sensitivity field of the imaging system (**Methods**). However, due to its scale invariance eMSOT only allows for quantifying blood $sO_2$ and not absolute blood volume, a goal that will be interrogated in future studies. Next steps further include the eMSOT validation with a larger pool of tissue physiology interrogations spanning from cancer, cardiovascular and diabetes research, relation of physiological phenotypes to metabolic and "-omic" outputs and in clinical application.

## ACKLOWLEDGEMENTS

Vasilis Ntziachristos acknowledges support from an ERC Advanced Investigator Award and the European Union project FAMOS (FP7 ICT, Contract 317744). The work of Stratis Tzoumas was supported by the DFG GRK 1371 grant. The authors would like to thank Elena Nasonova and Karin Radrich for assisting in the blood phantom preparation and Amir Rosenthal and Juan Aguirre for the valuable discussions.

**METHODS**

**Animal preparation and handling.** All procedures involving animal experiments were approved by the Government of Upper Bavaria. For the preparation of orthotopic 4T1 tumor models, 8 week old adult female athymic Nude-Foxn1 mice (Harlan, Germany) were orthotopically inoculated in the mammary pad with cell suspensions (0.5 million 4T1 cells (CRL-2539). Animals (n=8) were imaged *in-vivo* using MSOT after the tumors reached a suitable size. All imaging procedures were performed under anesthesia using 1.8% Isoflurane. In the $O_2$ challenge experiment, the mouse was initially breathing 100% $O_2$ and in the following medical air (20% $O_2$). During the $O_2$ Challenge, the mice were stabilized for a period of 10 minutes under each breathing condition before MSOT acquisition. For controlled mouse measurements (n=4), MSOT acquisition was performed on mice under gas anesthesia and breathing 100% $O_2$ or 20% $O_2$ by rectally inserting a capillary tube containing pig blood at 100% or 0% $sO_2$ oxygenation levels. Mice were sacrificed during MSOT imaging with an overdose of $CO_2$ or after MSOT acquisition by a Ketamine/Xylazine overdose. In the following the mice were stored at -80°C for further analysis.

4T1 cell line was acquired from ATCC (ATCC-CRL-2539, #5068892). The cells were authenticated by the ACTT by several analysis tests: Post-Freeze viability, Morphology, Mycoplasma contamination, post freeze cell growth, interspecies Determination; bacteria & fungal contamination. Additional mycoplasma contamination tests were also performed. For the animal studies no randomization, blinding or statistical methods were performed.

**MSOT.** Optoacoustic imaging was performed using a real-time whole body mouse imaging MSOT In Vision 256-TF (iThera-Medical GmbH, Munich, Germany). The system utilizes a cylindrically focused 256-element transducer array at 5MHz central frequency covering an angle of 270 degrees. The system acquires cross-sectional (transverse) images through the animal. The animals are placed onto a thin clear polyethylene membrane. The membrane separates the animals from a water bath, which is maintained at 34°C and is used for acoustic coupling and maintaining animal temperature while imaging. Image acquisition speed is at 10Hz[26]. Imaging was performed at 21 wavelengths from 700 nm to 900 nm with a step size of 10 nm and at 20 consecutive slices with a step size of 0.5 mm. Image reconstruction was performed using a model-based inversion algorithm [27] [28] with a non-negativity constraint imposed during inversion and with Tikhonov regularization.

**eMSOT method and $sO_2$ maps.** All optoacoustic images $P(\mathbf{r},\lambda)$ obtained over wavelength $\lambda$ were calibrated to correct for the intensity of laser power per pulse, and for the absorption of water surrounding the tissue. With $HbO_2$ and $Hb$ being the main tissue absorbers in the near-infrared, multispectral optoacoustic images can be related to the concentrations of oxy- and deoxy-hemoglobin through Eq. (2).

$$P(\mathbf{r},\lambda) = C(\mathbf{r}) \|\mathbf{\Phi}(\mathbf{r})\|_2 \frac{\Phi(\mathbf{r},\lambda)}{\|\mathbf{\Phi}(\mathbf{r})\|_2} \cdot (c_{HbO2}(\mathbf{r}) \cdot \varepsilon_{HbO2}(\lambda) + c_{Hb}(\mathbf{r}) \cdot \varepsilon_{Hb}(\lambda)). \quad (2)$$

In Eq. (2), $C(\mathbf{r})$ is a spatially varying scaling factor corresponding to the effects of the system's spatial sensitivity field and the Grüneisen coefficient, $\varepsilon_{HbO2}(\lambda)$ and $\varepsilon_{Hb}(\lambda)$ are the wavelength dependent molar extinction coefficients of oxygenated and deoxygenated hemoglobin, while $c_{HbO2}(\mathbf{r})$ and $c_{Hb}(\mathbf{r})$ the associated concentrations at a position $\mathbf{r}$. $\|\mathbf{\Phi}(\mathbf{r})\|_2$ is the norm of the light fluence across all wavelengths at a position $\mathbf{r}$, while $\Phi'(\mathbf{r},\lambda) = \Phi(\mathbf{r},\lambda)/\|\mathbf{\Phi}(\mathbf{r})\|_2$ is the normalized wavelength dependence of light fluence at a specific position (i.e. normalized spectrum of light fluence).

The space-only dependent factors $C(\mathbf{r})$ and $\|\mathbf{\Phi}(\mathbf{r})\|_2$ do not affect the estimation of blood $sO_2$ (Eq. (3)) which is calculated as a ratio once the relative concentrations of $HbO_2$ and $Hb$ are known (we define $c'_{HbO2}(\mathbf{r}) = C'(\mathbf{r}) \cdot c_{HbO2}(\mathbf{r})$ and $c'_{Hb}(\mathbf{r})=C'(\mathbf{r}) \cdot c_{Hb}(\mathbf{r})$, respectively, where $C'(\mathbf{r})$ is a common, space-only dependent scaling factor):

$$sO_2(\mathbf{r}) = \frac{c'_{HbO2}(\mathbf{r})}{c'_{HbO2}(\mathbf{r}) + c'_{Hb}(\mathbf{r})}. \quad (3)$$

For the accurate quantitative extraction of the relative values of $c'_{HbO2}(\mathbf{r})$ and $c'_{Hb}(\mathbf{r})$, accounting for, or estimating the wavelength dependence of the light fluence $\Phi'(\mathbf{r},\lambda)$ is further required.

**The *Eigenspectra* Model for Light Fluence.** eMSOT is based on the observation that the spectral patterns of light fluence present in tissue can be modeled as an affine function of only a few base spectra, independently of tissue depth and the specific distribution of optical properties of the tissue imaged. This hypothesis stems from the notion that the spectrum of light fluence is the result of the cumulative light absorption by hemoglobin; thus the spectrum of light fluence will always be related to the spectra of hemoglobin in a complex non-linear manner. This complex relation can be linearized using a data-driven approach, i.e. through the application of Principal Component Analysis (PCA) on a selected set of light fluence spectra.

The wavelength dependence of the light fluence was herein modeled as a superposition of a mean fluence spectrum $\Phi_M(\lambda)$ and a linear combination of a number of light fluence *Eigenspectra* $\Phi_i(\lambda)$. This model was derived by applying PCA on a training dataset comprised of a set of light fluence spectral patterns. Briefly, a training dataset was formed through the creation of multispectral light fluence simulations using the 1-D analytical solution of diffusion approximation for infinite media. A set of light fluence spectral patterns $\Phi_{z,ox}(\lambda)$ were computed for high physiological tissue optical properties ($\mu_a=0.3$ cm$^{-1}$, $\mu_s'=10$ cm$^{-1}$), tissue depths ranging from z=0 to z=1 cm with a step size of 0.143 mm (70 discrete depths in total) and for 21 different uniform background tissue oxygenations (ox $\in$ [0%, 5%, 10%, ..., 100%]). The so computed set of light fluence spectra $\Phi_{z,ox}(\lambda)$ was normalized ($\Phi'_{z,ox}(\lambda) = \Phi_{z,ox}(\lambda)/\|\Phi_{z,ox}\|_2$) and used in the following as training data in the context of PCA in order to create an affine, 3-dimensional model consisting of a mean fluence spectrum $\Phi_M(\lambda)$ and three *Eigenspectra* $\Phi_i(\lambda)$. PCA was used for offering a minimum square error property in capturing the spectral variability of light fluence in a linear manner. Three components were selected for providing a relatively simple model while also offering a small model error with respect to the training data-set (**Figure 1e**). The wavelength dependence of the light fluence $\Phi'(\mathbf{r},\lambda)$ at any arbitrary tissue position $\mathbf{r}$ can thus be modeled as a superposition of the mean fluence spectrum and three fluence *Eigenspectra* multiplied by appropriate scalar parameters $m_1$, $m_2$, and $m_3$, (hereby referred to as *Eigenfluence* parameters) as per Eq. (4):

$$\Phi'(\mathbf{r},\lambda) = \Phi_M(\lambda) + m_1\Phi_1(\lambda) + m_2\Phi_2(\lambda) + m_3\Phi_3(\lambda) \qquad (4)$$

The so created 3-dimensional affine forward model of the wavelength dependence of light fluence was tested with regard to light fluence spectral patterns produced in completely heterogeneous media with varying and randomly distributed optical properties and oxygenation values and demonstrated high accuracy (**Supplementary Fig. 1**). The forward model was further tested through *in-vivo* and *ex-vivo* light fluence measurements, obtained from controlled experiments (**Supplementary Fig. 2**).

Through simulations, it was observed that the values of the m$_2$ *Eigenfluence* parameter relate primarily to tissue depth and the average tissue optical properties. This trend was observed both in the case of tissue simulations with uniform optical properties (**Figure 1g**) as well as in complex and randomly created tissue simulations described in **Supplementary Note 1, 3**. Conversely, the values of the *Eigenfluence* parameters m$_1$ and m$_3$ relate both to tissue depth as well as to tissue background oxygenation. Specifically both m$_1$ and m$_3$ present a trend of increasing absolute values with depth and a sign that relates to background tissue oxygenation. These observations were confirmed with *in-vivo* and *ex-vivo* light fluence measurement experiments (**Supplementary Note 1**).

**Model Inversion.** Using the *Eigenspectra* model of light fluence, the blood sO$_2$ quantification problem at a position $\mathbf{r}$ formulates as the problem of estimating c'$_{HbO2}(\mathbf{r})$ and c'$_{Hb}(\mathbf{r})$ by minimizing $f(\mathbf{r};$ m$_1(\mathbf{r})$, m$_2(\mathbf{r})$, m$_3(\mathbf{r})$ c'$_{HbO2}(\mathbf{r})$, c'$_{Hb}(\mathbf{r}))$, for brevity noted $f(\mathbf{r})$, defined according to Eq, (5):

$$f(\mathbf{r}) = \left\| P(\mathbf{r},\lambda) - (\Phi_M(\lambda) + \sum_{i=1}^{3} m_i(\mathbf{r})\Phi_i(\lambda)) \cdot (c'_{HbO2}(\mathbf{r}) \cdot \varepsilon_{HbO2}(\lambda) + c'_{Hb}(\mathbf{r}) \cdot \varepsilon_{Hb}(\lambda)) \right\|_2 . \qquad (5)$$

The solution for the 5 unknowns (namely the 3 light fluence model parameters, $m_{1...3}(\mathbf{r})$ and the relative blood concentrations c'$_{HbO2}(\mathbf{r})$ and c'$_{Hb}(\mathbf{r})$) can be obtained using a non-linear optimization algorithm and at least 5 excitation wavelengths. The relative blood concentrations c'$_{HbO2}(\mathbf{r})$ and c'$_{Hb}(\mathbf{r})$ are proportional to the actual ones (c$_{HbO2}(\mathbf{r})$ and c$_{Hb}(\mathbf{r})$) with regard to a common scaling factor. However, as stated before, this fact does not affect the computation of sO$_2$.

The minimization problem defined by Eq. (5) is ill-posed and may converge to a wrong solution unless properly constrained. For achieving inversion stability and accurate sO$_2$ estimation results, the cost function $f$ of Eq. (5) is simultaneously minimized in a set of grid points placed in the image domain (**Figure 1i**), where three independent constraints are further imposed to the *Eigenfluence* parameters. These constraints correspond to the relation of the *Eigenfluence* parameters between neighbor grid points and to the allowed search space for the *Eigenfluence* parameters:

(i) Since the values of the second *Eigenfluence* parameter m$_2$ present a consistent trend of reduction with tissue depth observed both in the case of uniform tissue simulations (see **Fig. 1g**) as well as in simulations with random structures, m$_2$ is constrained to obtain smaller values in the case of grid points placed deeper into tissue.

(ii) Since the light fluence is bound to vary smoothly in space due to the nature of diffuse light propagation, large variations of the *Eigenfluence* parameters m$_1$, and m$_3$ between neighbor pixels are penalized. This is achieved through the incorporation of appropriate Lagrange multipliers $\lambda_i$ to the cost function for constraining the variation of the model parameters (Eq. (6)). The values of the Lagrange multipliers were selected using cross-validation on simulated data-sets (**Supplementary Note 2**).

(iii) Since the values of m$_1$ and m$_3$ are strongly dependent on background tissue oxygenation, an initial less accurate estimation of tissue sO$_2$ can be effectively used to reduce the total search-space to a constrained relevant sub-space. The limits of search space for the *Eigenfluence* parameters m$_1$ and m$_3$ corresponding to each grid point are identified in a preprocessing step as analytically described in **Supplementary Note 2**.

Assuming a circular grid of P arcs and L radial lines (see **Suppl. Fig. 3**) with a total of P x L points $\mathbf{r}_{p,l}$, and let the vector $\mathbf{m_i}$ =[m$_i(\mathbf{r}_{1,1})$, m$_i(\mathbf{r}_{1,2})$, ..., m$_i(\mathbf{r}_{1,L})$, m$_i(\mathbf{r}_{2,1})$, ..., m$_i(\mathbf{r}_{p,l})$,...,m$_i(\mathbf{r}_{P,L})$] correspond to the values of the light fluence parameter $i$ ($i=1...3$) over all such points, the new inverse problem is defined as the minimization of cost function $f_{grid}$ defined in Eq. (6) under the constraints defined in Eq. (7).

$$f_{grid} = \sum_i f(\mathbf{r}_i) + \lambda_1 \|\mathbf{Wm}_1\|_2 + \lambda_3 \|\mathbf{Wm}_3\|_2 \quad (6)$$

$$\lim_2^{min} < m_2(\mathbf{r}_k) < \lim_2^{max}, \forall \mathbf{k},$$
$$m_2(\mathbf{r}_{p+1,l}) < m_2(\mathbf{r}_{p,l}), m_2(\mathbf{r}_{p+1,l+1}) < m_2(\mathbf{r}_{p,l}), m_2(\mathbf{r}_{p+1,l-1}) < m_2(\mathbf{r}_{p,l}), \forall p,l,$$
$$\lim_{i,\mathbf{r}_k}^{min} < m_i(\mathbf{r}_k) < \lim_{i,\mathbf{r}_k}^{max}, \forall \mathbf{k}, i=1,3, \quad (7)$$
$$c'_{HbO_2}(\mathbf{r}_k) \geq 0, \forall \mathbf{k},$$
$$c'_{Hb}(\mathbf{r}_k) \geq 0, \forall \mathbf{k},$$

In Eq. (6), **W** is the weighted connectivity matrix corresponding to grid of points assumed (**Supplementary Note 2**). Each matrix element corresponds to a pair of grid points $\mathbf{r}_{p1,l1}$ $\mathbf{r}_{p2,l2}$ and is zero if the points are not directly connected or inverse proportional to their distance ($w(\mathbf{r}_{p1,l1}$ $\mathbf{r}_{p2,l2}) = 1/\|\mathbf{r}_{p1,l1} - \mathbf{r}_{p2,l2}\|_2$) if the points are connected. The inverse problem defined by Eq. (6), (7) was hereby solved through the utilization of sequential quadratic programming algorithm of MATLAB toolbox.

**Fluence correction and sO₂ quantification.** The minimization of cost function $f_{grid}$ (Eq. (6)) under the constraints of Eq. (7) yields an estimate of $m_i(\mathbf{r})$ for each *Eigenfluence* parameter *i* and each grid point **r**. The *Eigenfluence* parameters in the convex hull of the grid are in the following estimated by means of cubic interpolation. We note that due to the nature of diffuse light propagation the *Eigenfluence* parameters are expected to vary smoothly in tissue and thus their interpolation is not expected to introduce large errors in the result (see **Supplementary Note 3**). The wavelength dependence of light fluence is computed for each pixel within the convex hull of the grid as in $\Phi'(\mathbf{r},\lambda) = \Phi_M(\lambda) + m_1(\mathbf{r})\Phi_1(\lambda) + m_2(\mathbf{r})\Phi_2(\lambda) + m_3(\mathbf{r})\Phi_3(\lambda)$, where $\Phi_i(\lambda)$ is the *i*th fluence *Eigenspectrum*. Finally, a spectrally-corrected eMSOT image is obtained after diving the original image $P(\mathbf{r},\lambda)$ with the normalized wavelength dependent light fluence $\Phi'(\mathbf{r},\lambda)$ at each position **r** and wavelength $\lambda$, i.e. $P^{eMSOT}(\mathbf{r},\lambda) = P(\mathbf{r},\lambda)/\Phi'(\mathbf{r},\lambda)$. Blood sO₂ is computed for each pixel of $P^{eMSOT}(\mathbf{r},\lambda)$ image independently through nonnegative constrained least squares fitting with the spectra of oxygenated and deoxygenated hemoglobin. Thus the eMSOT blood sO₂ maps retain the original resolution of the MSOT imaging system.

We note that both the *Eigenspectra* model and the inversion scheme were hereby optimized for the application of small animal imaging. The *Eigenspectra* model was trained for a maximum depth of 1 cm and the inversion scheme was designed with respect to the same tissue depth and optical properties within the physiological range (**Supplementary Note 2, 3**).

**Blood Phantom Preparation.** For validating the accuracy of eMSOT in quantifying blood oxygenation in deep tissue, we prepared tissue mimicking phantoms, containing blood at known oxygenations levels. Specifically, for simulating tissue background, 2cm –diameter cylindrical solid phantoms were created by using 1.5% Agarose Type I, Sigma-Aldrich (solidifying in <37º), 2% intralipid and 3-5% freshly extracted pig blood diluted in NaCl. Different blood oxygenation levels were achieved by diluting oxygen in whole blood (oxygenation process) or by mixing the blood with different amounts of Sodium Dithionite ($Na_2O_4S_2$) (deoxygenation process) [29]. The levels blood oxygenation were monitored using a Bloodgas Analyzer (Eschweiler Gmbh & Co. KG, Kiel Germany).

**Cryoslicing color imaging and H&E staining of tumor tissues.** After MSOT acquisition, a subset of the mice bearing 4T1 mammarian tumors (n=4) were sacrificed and examined for tumor and tissue anatomy. Mice were embedded in an optimal cutting temperature compound (Sakura Finetek Europe BV, Zoeterwonde, NL) and frozen at -80°C. In the following the mice were sliced at an orientation similar to the MSOT imaging and color photographs were recorded. The cryoslicing imaging system is based on a cryotome (CM 1950, Leica Microsystems, Wetzlar, Germany), fitted with CCD-based detection camera. During this process, 10 µm slices throughout the whole tumor volume were collected for further histological analysis.

Several slides per tumor were subjected to H&E staining and imaging. The slides containing 10µm cryo-sections were first pre-fixed in 4% PFA (Santa Cruz Biotechnology Inc., Dallas, Texas, USA). Then, they were rinsed with distilled water and incubated 30 seconds with Haemotoxylin acide by Meyer (Carl Roth, Karlsruhe, Germany) to stain the cell nuclei. The slides were then rinsed in tap water again before incubation for 1 second in Eosin G (Carl Roth, Karlsruhe, Germany) to stain cellular cytoplasm. After rinsing in distilled water, the slides were dehydrated in 70%, 94% and 100% ethanol and incubated for 5 minutes in Xylene (Carl Roth, Karlsruhe, Germany) before being cover slipped with Rotimount (Carl Roth, Karlsruhe, Germany) cover media. Representative slides were observed using Zeiss Axio Imager M2 microscope with AxioCam 105 Color, and pictures were then processed using a motorized stitching Zen Imaging Software (Carl Zeiss Microscopes GmbH, Jena, Germany).

**Pimonidazole Staining of tumor tissues.** A subset of the tumor-bearing mice (n=4) was examined for functional characteristics of the tumors by Pimonidazole histological staining. The hypoxia marker Pimonidazole (Hypoxyprobe, catalog #HP6-100 kit, Burlington, MA, USA) was injected i.p. at 100 mg/kg body weight in a volume of 0.1 ml saline ≈1.5h before tumor excision, and the perfusion marker Hoechst 33342 (Sigma, Deisenhofen, Germany) was administered i.v. at 15 mg/kg body weight in a volume of 0.1 ml saline 1min before the tumor-bearing mice were sacrificed. The tumors were excised immediately after the animals were sacrificed. The orientation of the tumors with respect to the mouse body was retained. 8 µm cryosections were sliced throughout the tumor. The cryosections were fixed in cold (4°C) acetone, air dried and rehydrated in PBS before staining. Pimonidazole was stained with the FITC-labelled anti-Pimonidazole antibody (Hypoxyprobe, Burlington, MA, USA) diluted 1:50 in primary antibody diluent (PAD, Serotec, Oxford, U.K.) by incubating for 1h at 37°C in the dark.

## SUPPLEMENTAL MATERIAL

**Supplementary Note 1: Numerical and experimental validation of the *Eigenspectra* model of light fluence (forward model validation).**

For validating the accuracy of the *Eigenspectra* model for light fluence ($\Phi_M(\lambda)$, $\Phi_1(\lambda)$, $\Phi_2(\lambda)$, $\Phi_3(\lambda)$) over light fluence spectra created in arbitrary tissues, we created simulations of the absorbed energy density of arbitrary tissues at different wavelengths (700 nm to 900 nm with a step of 10 nm), using light propagation models. Assuming a circular structure of 1 cm radius, random maps of optical absorption ($\mu_\alpha(\mathbf{r})$) and reduced scattering coefficient ($\mu_s'(\mathbf{r})$) were formed (**Supplementary Fig. 1a** and **b**, respectively), the values of which follow a Gaussian distribution ($\mu_\alpha \in N(\mu_\alpha^{mean}, \mu_\alpha^{std})$ where $\mu_\alpha^{mean} \in [0.07, 0.3]$ cm$^{-1}$ and $\mu_\alpha^{std}=0.1$ cm$^{-1}$, $\mu_s' \in N(\mu_s^{mean}, \mu_s^{std})$) where $\mu_s^{mean} \in [7, 11]$ cm$^{-1}$ and $\mu_s^{std}=3$ cm$^{-1}$). The so created absorption maps ($\mu_\alpha(\mathbf{r})$) correspond to tissue absorption at an excitation wavelength of 800 nm (isosbestic point of hemoglobin). The absorption maps for different excitation wavelengths are computed based on the one at 800 nm and the absorption spectra of oxy- and deoxy-hemoglobin. The relative amount of oxy- versus deoxy-hemoglobin at each position $\mathbf{r}$ is defined by a random map of tissue blood oxygenation (**Supplementary Fig. 1c**). Different blood sO$_2$ maps were simulated (one example presented in **Supplementary Fig. 1c**) with spatially varying random oxygenation values, and with an average tissue oxygenation varying from ~10% to 90% and a standard deviation of 30%.

The multispectral absorption and scattering maps were employed in a 2D finite-element-method (FEM) solution of the diffusion equation (DE)[1] to simulate multispectral optoacoustic data-sets (i.e. multi-wavelength absorbed energy density) of tissue with arbitrary structure, optical properties and oxygenation. One such example is shown in **Supplementary Figure 1d** for a single wavelength. From these datasets, the normalized wavelength dependent light fluence $\Phi'(\mathbf{r},\lambda)=\Phi(\mathbf{r},\lambda)/\|\Phi(\mathbf{r})\|_2$ was calculated for each position $\mathbf{r}$ in the image. The residual value obtained after comparing the simulated fluence spectra $\Phi'(\mathbf{r})$ to their approximation using the basis functions of the *Eigenspectra* model ($\Phi'_{Model}(\mathbf{r})$) was computed (res $=\|\Phi'(\mathbf{r})-\Phi'_{Model}(\mathbf{r})\|_2/\|\Phi'(\mathbf{r})\|_2$) for each pixel in the image $\mathbf{r}$ and statistics of this value are presented in **Supplementary Figure 1e.** Statistics correspond to all pixels of 21 simulations per mean oxygenation, corresponding to different mean optical absorption and scattering (231 simulations in total). **Supplementary Figure 1f** further plots the error of the forward model in the sO$_2$ estimation (i.e. the error propagated in sO$_2$ estimation due to the approximation of $\Phi'(\mathbf{r},\lambda)$ with $\Phi'_{Model}(\mathbf{r},\lambda)$).

The *Eigenspectra* forward model was tested with 231 simulations of high (**Suppl. Fig. 1a-c**) and 231 simulations of low spatial variation of optical properties (**Suppl. Fig. 1g**) and oxygenation. Moreover the forward model was tested in simulations of blob-like features (representing organs) and vessel-like structures (**Suppl. Fig. 1h**). In this case, the blob-like structures correspond to $\mu_\alpha = 0.3$ cm$^{-1}$, the background to $\mu_\alpha = 0.1$ cm$^{-1}$ and the vessel like structures to $\mu_\alpha = 5.4$ cm$^{-1}$ and $\mu_s' = 16$ cm$^{-1}$. The $\mu_s'$ and sO$_2$ maps corresponding to the background followed a random distribution as previously described and the sO$_2$ of the vessel-like structure was retained uniform and 25% higher than the mean oxygenation of the background. Statistics on the fitting residual of the forward model on the simulations of **Supplementary Figure 1a, g, h** are presented in **Supplementary Figure 1e, j, k**, respectively. We observed a small error in the forward model independently of tissue structure and the variations of optical properties and tissue oxygenation.

To assess the potential influence of parameters not included in the model such as the absorption of melanin and the wavelength dependence of scattering we further created simulations containing a strongly absorbing melanin component at the tissue surface ($\mu_\alpha = 2.5$ cm$^{-1}$) and an exponentially decaying scattering coefficient ($\mu_s'= 18.9(\lambda/500)^{-0.6}$ cm$^{-1}$) that corresponds to whole blood measurements[2]; an example presented in **Supplementary Figure 1i**. The assumed optical properties were again following a normal distribution with $\mu_\alpha \in N(\mu_\alpha^{mean}, \mu_\alpha^{std})$ where $\mu_\alpha^{mean} \in [0.07, 0.3]$ cm$^{-1}$ and $\mu_\alpha^{std}=0.1$ cm$^{-1}$, $\mu_s' \in N(\mu_s^{mean}, \mu_s^{std})$) where $\mu_s^{mean} \in [7, 11]$ and $\mu_s^{std} =3$ cm$^{-1}$ (21 simulations per mean oxygenation, 231 simulations in total). Similar to the absorption maps, the so created scattering maps $\mu_s'(\mathbf{r})$ correspond to tissue scattering at an excitation wavelength of 800 nm. The scattering maps for different excitation wavelengths are computed based on the one at 800 nm and the exponentially decaying curve of the scattering coefficient. In this case the fitting residual of the forward model is increased (**Suppl. Fig. 1 l**) but is still preserved in relatively low levels indicating that the model retains accuracy despite the simplifying assumptions in its creation.

The accuracy of the forward model in the ballistic regime was tested using Monte Carlo simulations[3] of multi-layered tissue (**Suppl. Fig. 1 m**). Four different tissue layers were assumed with different oxygenation levels and optical properties. In this case the fitting residual of the forward model is similar to the one when using the diffusion approximation: 0.61±0.22%.

The graphs indicate a small model error, supporting the hypothesis that a simple affine model with only three *Eigenspectra* can capture the spectral variability of $\Phi'(\mathbf{r},\lambda)$ in complex tissue structures, independently of the distribution of the optical properties. We hereby note that the error in oxygenation depicted in **Supplementary Figure 1f** is just indicative of the model accuracy (error of the forward model) and does not relate to the actual blood sO$_2$ estimates that can be obtained through this procedure by solving the inverse problem (estimation error of the inverse problem).

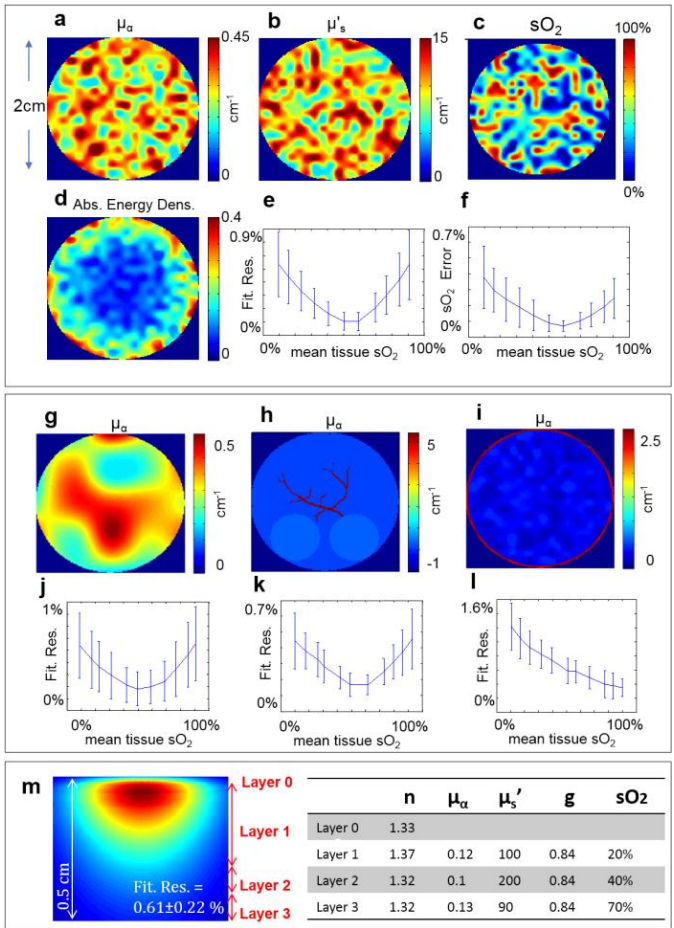

**Supplementary Figure 1. Numerical validation of the *Eigenspectra* model of light fluence in tissue simulations of arbitrary structures.** (**a,b**) Random spatial map of (**a**) $\mu_a$ at 800 nm and (**b**) $\mu_s'$ with random, normally distributed values. (**c**) Random spatial map of $sO_2$. (**d**) Example of multi-wavelength absorbed energy density simulation, at wavelength 800 nm, created using the FEM DE light propagation model. (**e**) Statistics (mean and standard deviation - errorbar) on the fitting residual of the *Eigenspectra* model computed from all pixels of each simulated multispectral dataset. (**f**) Error propagated to $sO_2$ estimation due to the fluence approximation using the *Eigenspectra* model. (**g-i**) Tissue simulations of low spatial variation of optical properties (**g**), partially uniform optical properties with highly absorbing vessel like structures (**h**) and cases of high melanin absorption in the tissue surface as well as wavelength dependent scattering (**i**). (**j-l**) Statistics of the fitting residual of the forward model corresponding to **g-i**, respectively. (**m**) Monte Carlo simulations of the wavelength dependent light fluence (fluence in one wavelength is presented) in the ballistic and semi-ballistic regime, assuming semi-uniform multi-layered tissue; Layers are highlighted with red arrows and their optical properties are summarized in the enclosed table.

To experimentally investigate the validity of the *Eigenspectra* model of tissue light fluence we obtained measurements from small animals *in-vivo* and *post-mortem*. We measured the light fluence in tissue by inserting a reference chromophore with well characterized spectrum within tissue. Specifically, a capillary tube was rectally inserted into an anesthetized CD1 mouse and the animal was imaged in the lower abdominal area *in-vivo* using the MSOT system. The capillary tube was filled with black India ink, the spectrum of which was previously measured in the photospectrometer. The animal was imaged *in-vivo* under 100% $O_2$ breathing and *ex-vivo*. These two different physiological conditions were employed in order to investigate the influence of the average background tissue oxygenation on the spectrum of the light fluence.

The per-wavelength image intensity at the region of the ink insertion (i.e. the optoacoustic measured spectrum which corresponds to the multiplication of the local absorption with the local light fluence) was elementwise divided by the actual absorption spectrum of ink. The resulting spectrum after division corresponds to the wavelength dependence of the local light fluence. The measured light fluence spectrum computed in this way was fitted to the *Eigenspectra* model and the two curves and the fitting residual are presented in **Supplementary Figure 2**.

**Supplementary Figure 2a** presents a single wavelength optoacoustic image of the mouse in the abdominal area. The area where the light fluence is measured is indicated with a red circle. **Supplementary Figure 2b** presents the spectrum of the experimentally measured light fluence (black curves) and the fitting result using the *Eigenspectra* model in the case of *in-vivo* (blue curve) and *post-mortem* imaging (red curve). The low fitting residuals indicate good agreement of the model with experimental reality. **Supplementary Figure 2c** presents the decomposition of the two fitted light fluence spectra as a linear combination of the mean fluence spectrum and the three *Eigenspectra*. While the first and the third *Eigenspectra* components change dramatically with respect to the two different tissue oxygenation states, the second component that corresponds to tissue depth remains relatively unchanged. Moreover the values of the $m_1$ parameter obtained after fitting were positive in the *post-mortem* case and negative in the *in vivo* case, an observation that is in accordance with the dependence of $m_1$ on background tissue oxygenation as presented in **Figure 1f**. This observation was confirmed by performing the same experiment in 2 more animals. Overall, the low fitting residual even in the case of experimental data obtained *in-vivo* indicates good agreement between theory and experimental reality.

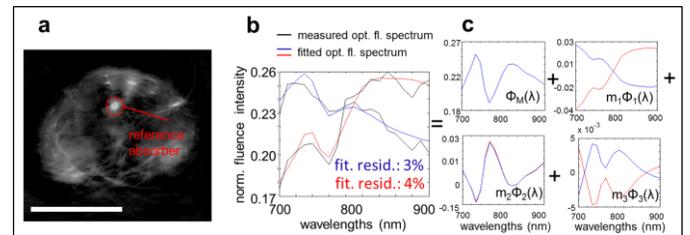

**Supplementary Figure 2. Validation of the *Eigenspectra* model in the case of tissue data obtained *in-vivo*.** (**a**) MSOT image (one wavelength presented) of a CD1 mouse imaged in the abdominal region with a capillary tube containing a reference absorber inserted in the lower abdominal area (red circle). Scale bar, 1 cm. (**b**) Comparison of the measured spectrum of light fluence in the area of absorber insertion (black curves) with the fitted result using the 3-dimensional *Eigenspectra* model in the case of *in-vivo* imaging (blue curve) and *post-mortem* imaging (red curve). (**c**) The two light fluence spectra can be decomposed in a linear combination of spectra $\Phi_M(\lambda)$, $m_1\Phi_1(\lambda)$, $m_2\Phi_2(\lambda)$ and $m_3\Phi_3(\lambda)$.

**Supplementary Note 2: Constrained inversion**

The spatial characteristics of light fluence were exploited for overcoming the ill-posed nature of the optimization problem defined by Eq. (5). In contrary to tissue absorption which can vary arbitrarily, the light fluence is bound to vary smoothly in space due to the nature of diffuse light propagation. In the context of the *Eigenspectra* model inversion, such *a priori* information can be incorporated by attempting simultaneous inversion on a grid of points in the image domain (an example of such a grid is shown in **Suppl. Fig. 3a**), and penalizing large variations of the *Eigenfluence parameters* between neighbor pixels. A matrix **W** representing a weighted non-directed graph (**Supplementary Figure 3b**) is defined based on the assumed grid, where neighbor grid points are connected with weighted vertexes. We penalize large spatial variations of light fluence by incorporating the norm of the variation of the *Eigenfluence* parameters $m_1$ and $m_3$ in the minimization function using Lagrange multipliers: Assuming a circular grid of P arcs and L radial lines (**Supplementary Figure 3b**), with a total of P x L points $\mathbf{r}_{p,l}$, and let the vector $\mathbf{m}_i = [m_i(\mathbf{r}_{1,1}), m_i(\mathbf{r}_{1,2}), \ldots, m_i(\mathbf{r}_{1,L}), m_i(\mathbf{r}_{2,1}), \ldots, m_i(\mathbf{r}_{p,l}), \ldots, m_i(\mathbf{r}_{P,L})]$ correspond to the values of the *Eigenfluence* parameter $i$ ($i=1\ldots3$) over all such points, the new cost function $f_{grid}$ is defined by Eq. (2.1):

$$f_{grid} = \sum_i f(\mathbf{r}_i) + \lambda_1 \| \mathbf{W}\mathbf{m}_1 \|_2 + \lambda_3 \| \mathbf{W}\mathbf{m}_3 \|_2, \quad (2.1)$$

where $f$ is the cost function defined in Eq. (5) and **W** is the weighted connectivity matrix corresponding to grid of points assumed. Each matrix element corresponds to a pair of grid points $\mathbf{r}_{p1,l1}$ $\mathbf{r}_{p2,l2}$ and is zero if the points are not directly connected or inverse proportional to their distance ($w(\mathbf{r}_{p1,l1} \mathbf{r}_{p2,l2}) = 1/\| \mathbf{r}_{p1,l1} - \mathbf{r}_{p2,l2} \|_2$) if the points are connected.

The values of the Lagrange multipliers $\lambda_1$ and $\lambda_3$ were selected using cross-validation on simulated data-sets with finely granulated structures (**Supplementary Figure 1a-c**). We did not observe high sensitivity of the result obtained to small changes of the Lagrange multipliers. The same values for the Lagrange multipliers were used for all simulated and experimental data presented in the work.

An alternative spatial fluence constraint is applied in the case of the second *Eigenfluence* parameter $m_2$. Through simulations of uniform optical properties as well as simulations with randomly varying optical properties it was observed that the values of $m_2$ are strongly and consistently associated with tissue depth, obtaining lower values in deeper tissue areas. Through the definition of an additional directed graph based on the assumed gird (**Supplementary Figure 3c**) the value of $m_2$ at a certain grid point was enforced to obtain larger values than the ones of its direct neighbors placed deeper in tissue:

$$m_2(\mathbf{r}_{p+1,l}) < m_2(\mathbf{r}_{p,l}), m_2(\mathbf{r}_{p+1,l+1}) < m_2(\mathbf{r}_{p,l}), m_2(\mathbf{r}_{p+1,l-1}) < m_2(\mathbf{r}_{p,l}), \forall p,l, \quad (2.2)$$

For further enhancing the inversion stability, additional constraints were imposed to the *Eigenfluence* parameters that relate to both depth and background tissue oxygenation (i.e. $m_1$ and $m_3$) based on a first approximate estimate of tissue blood oxygenation. By performing linear spectral unmixing on the raw multispectral optoacoustic images $P(\mathbf{r},\lambda)$ a first estimation map of blood $sO_2$ levels can be obtained. It is noted that this $sO_2$ map is incrementally erroneous with tissue depth, however it can serve as a first approximation for constraining the total search-space for $m_1$ and $m_3$ to a more relevant sub-space. Using the so created $sO_2$ map (**Suppl. Figure 3d**) and by assuming uniform tissue optical properties (i.e. $\mu_a = 0.3$ cm$^{-1}$ at 800 nm and $\mu_s'=10$ cm$^{-1}$) a light fluence map is simulated using a FEM of the DA. By fitting the simulated light fluence spectra $\Phi'(\mathbf{r},\lambda)$ to the *Eigenspectra* model, prior estimates of all model parameters $\acute{m}_1(\mathbf{r})$, $\acute{m}_2(\mathbf{r})$ and $\acute{m}_3(\mathbf{r})$ can be obtained for each grid point **r**. A map of $\acute{m}_1$ corresponding to the $sO_2$ map of **Suppl. Figure 3d** is presented in **Suppl. Figure 3e** while the values of $\acute{m}_1(\mathbf{r})$ for all positions **r** in one radial line of the gird in **Suppl. Figure 3a** are presented in **Suppl. Figure 3f** (blue line).

The optimization problem of Eq. (2.1) is solved, with the values of $m_1(\mathbf{r})$ and $m_3(\mathbf{r})$ constrained to lie within a region surrounding the initial prior estimate $\acute{m}_i(\mathbf{r})$ (blue vertical lines in **Suppl. Figure 3f**):

$$\lim_{i,\mathbf{r}_k}^{min} < m_i(\mathbf{r}_k) < \lim_{i,\mathbf{r}_k}^{max}, \forall \mathbf{k}, i=1,3. \quad (2.3)$$

The limits of the allowed search space ($\lim_{i,\mathbf{r}_k}^{min}$, $\lim_{i,\mathbf{r}_k}^{max}$) were selected *ad hoc* as a function of the prior *Eigenfluence* values $\acute{m}_i$ and tissue depth, through the comparison of the prior and the real *Eigenfluence* parameters computed in tissue simulations of varying (uniform) optical properties ($\mu_a \in [0.1$-$0.3]$cm$^{-1}$ at 800 nm, $\mu_s'=10$ cm$^{-1}$) and all uniform oxygenation levels. It is noted that the allowed search space is incrementally larger with tissue depth since in deep tissue the original $sO_2$ estimates (and thus the *Eigenfluence* priors) usually deviate significantly from the true values. **Supplementary Figure 3f** presents an example of constrained inversion corresponding to a radial grid line of the simulation of **Supplementary Figure 3a**: The blue line indicates the prior $\acute{m}_1(\mathbf{r})$ across the grid pixels, the blue vertical lines indicate the limits of search space, the green line indicates the actual $m_1(\mathbf{r})$ values of the grid points and the red line the estimated ones after nonlinear optimization. The same function for computing the limits ($\lim_{i,\mathbf{r}_k}^{min}$, $\lim_{i,\mathbf{r}_k}^{max}$) as a function of the prior $\acute{m}_i$ estimate and tissue depth was used for all simulated and experimental data presented in the work.

We note that this constraint (identified through trends in uniform tissue data) may not always be exact in data of complex structures of optical properties and oxygenation; thus excluding in certain cases the optimal solution from the allowed search space. Despite this, the evaluation of **Supplementary Note 4** indicated that the enforcement of this constraint typically leads to a solution close to the optimal one even in such cases, while it minimizes the possibility or an irrelevant convergence in all cases; sacrificing thus accuracy for robustness.

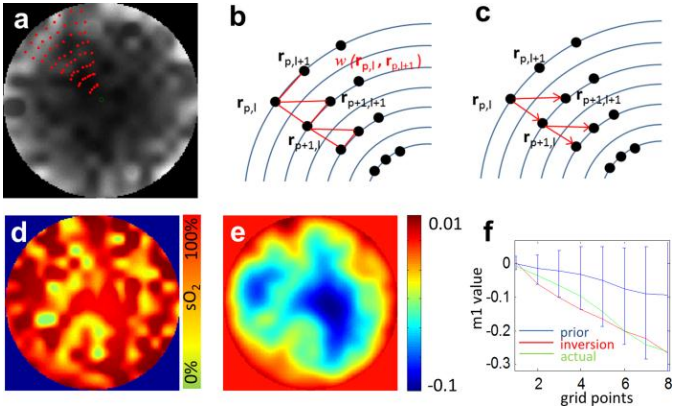

**Supplementary Figure 3. Incorporation of constraints on the values of the *Eigenfluence* parameters $m_1$, $m_2$ and $m_3$.** (**a**) Inversion is performed simultaneously on a grid of points in the image domain. (**b**) A non-directed weighted graph on the grid of points penalizes large variations of the *Eigenfluence* parameters between neighbor points. The penalization is inversely proportional to the distance $w$ between the grid points. (**c**) A directed graph on the grid of points enforces a decrease on the values of $m_2$ with depth. (**d-f**) An initial approximation of tissue blood oxygenation is obtained using nonnegative constrained least squares fitting (**d**) and used for obtaining a prior estimate of $m_1$ (**e**) and $m_3$ model parameters. These prior estimates are used for constraining the total search space. (**f**) Prior $\acute{m}_1$ estimate (blue line), limits of the search space (blue vertical lines), actual $m_1$ values (green line) and $m_1$ values estimated after optimization (red line) for a radial line of the grid presented in (**a**).

**Supplementary Note 3: Numerical validation of eMSOT.**

For investigating the ability of eMSOT to obtain accurate quantitative estimates of tissue blood oxygenation we validated its performance using numerical simulations of multi-wavelength absorbed energy density. The absorbed energy density simulations were formed as described in **Supplementary Note 1** using random or semi-random maps of absorption, scattering coefficient and blood oxygenation. A large validation data-set of more than 2000 different simulations was employed. The optical properties and $sO_2$ maps followed a random spatial variation with different structural characteristics ranging from finely granulated to smoothly varying structures (**Suppl. Fig. 4a**) as well as highly absorbing vascular structures with an absorption coefficient ranging from 1 to 6 times larger than the mean tissue background (**Suppl. Fig. 4a right low**). In each case the mean tissue optical properties varied from low to high tissue absorption and scattering (**Suppl. Fig. 4b**) in the physiological range ($\mu_a^{mean}$= 0.07 0.1, 0.15, 0.2, 0.25, 0.3 cm$^{-1}$ at 800 nm and $\mu_s^{mean}$ = 7, 9, 11 cm$^{-1}$). For each combination of $\mu_a^{mean}$, $\mu_s^{mean}$, different random blood $sO_2$ maps were assumed ranging from a mean tissue oxygenation of 10% to 90%. Random Gaussian noise with energy varying from 2.5% to 4.5% of the original energy of the spectra in each pixel was further superimposed.

**Supplementary Figure 4c** presents a simulated multispectral optoacoustic image (one wavelength presented) after incorporating the optical property maps in a FEM solution of the diffusion equation. An arc-shaped grid of 50 points is applied in the upper-left part of the simulation for the application of the eMSOT method. The parameters of inversion and the constraints employed were the same with the ones used for analyzing the *in-vivo* datasets and are analytically described in **Methods** and **Supplementary Note 2**. An example of the original (green) and the noisy spectrum (blue) corresponding to a pixel of **Suppl. Fig. 4c** with 4.5% superimposed random noise is visualized in **Suppl. Fig. 4d**. **Supplementary Figure 4e-g** present the recovered maps of the *Eigenfluence* parameters $m_1$, $m_2$ and $m_3$ after inversion and interpolation in the convex hull of the grid. **Suppl. Fig. 4h-j** present the $sO_2$ estimation using nonnegative constrained least squares fitting with the spectra of oxy- and deoxy-hemoglobin on the original simulation (**h**), eMSOT $sO_2$ estimation (**i**), as well as the actual simulated $sO_2$ map (**j**). **Supplementary Figure 4k** presents the corresponding errors in $sO_2$ estimation of the eMSOT method (blue points) and linear unmixing (red points) in all pixels of the analyzed area, sorted per depth. The $sO_2$ estimation error maps in the whole analyzed area were used for statistically evaluating the eMSOT performance.

Upon evaluation of the method on a set of more than 2000 randomly created simulations, we observe that in the physiological range of mean tissue oxygenation between 30% and 80% the mean $sO_2$ estimation error ranges from 2.4% to 3.4% depending on the levels of random noise, while in ~97% of the cases the $sO_2$ error did not exceed 10% (**Supplementary Table 1**). We did not observe dramatic performance differences between different mean optical properties or different structures of the optical properties. We further did not observe significant performance degradation with high levels of superimposed noise indicating that the inversion scheme is rather robust to noise. The largest errors were observed in the case of less than 30% mean tissue oxygenation. In this case the mean $sO_2$ error was 5% and in ~97% of the cases the error was less than 15%. The results of the statistical evaluation of the method over all simulations tested are analytically presented in **Supplementary Table 1**.

**Supplementary Figure 4l** presents the mean $sO_2$ error of linear unmixing and eMSOT corresponding to each simulated data-set tested, while **Supplementary Figure 4m** present the histogram of the mean $sO_2$ error corresponding to all simulations. In 88% of all cases tested the eMSOT method offered a lower mean estimation error than conventional linear unmixing. In the rest 12% of the cases linear unmixing offered a better estimation, but the mean $sO_2$ errors were comparable and both were lower than 8%. Finally, **Supplementary Figure 4n** presents a histogram of the relative $sO_2$ error yielded by linear unmixing over eMSOT for all simulated data-sets tested and for simulated tissue depths>5mm; indicating that eMSOT typically offered 3 to 8-fold enhanced $sO_2$ estimation accuracy in deep tissue.

The statistical evaluation of **Suppl. Table 1** corresponds to the application of a circular grid of an angle step of $\pi/20$ rads and a radial step of 0.14 cm. The effect of the grid density in the $sO_2$ estimation accuracy was further tested through the application of different grid densities spanning from 12, 30, 49 and 108 points in a $\pi/4$ disk area; the results are summarized in **Suppl. Table 2**. We observed that the $sO_2$ estimation accuracy does not increase dramatically with an increased grid density due to the smooth spatial variations of light fluence in tissue.

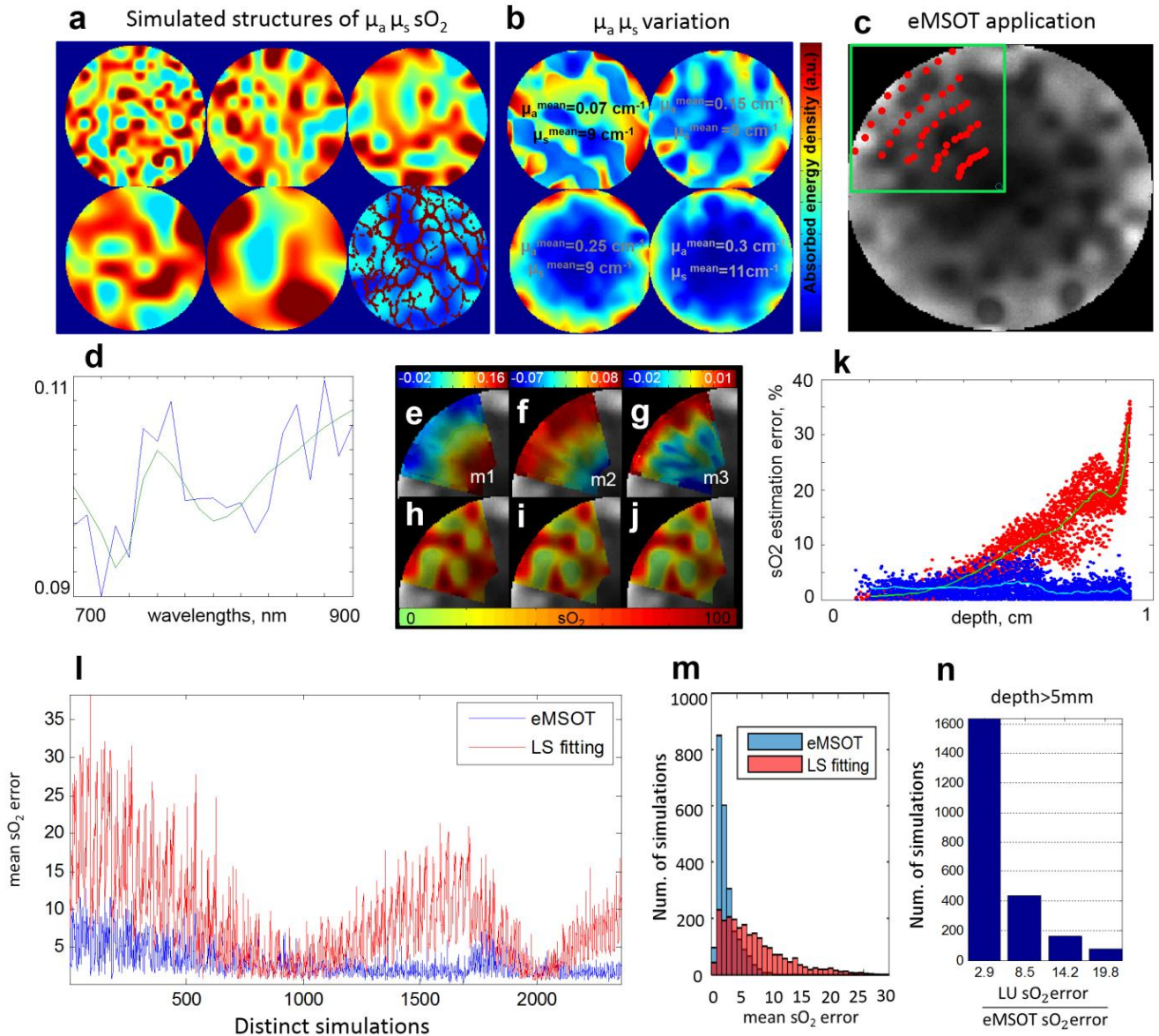

**Supplementary Figure 4. Numerical validation of eMSOT in simulations of arbitrarily structured tissues.** (**a**) Examples of the assumed random maps of optical absorption, optical scattering and $sO_2$ varying from finely granulated to smoothly varying structures and vessel-like patters. The combination of these maps was used to simulate the absorbed energy density of complex tissue using a light propagation model. (**b**) The absorbed energy densities were formed with varying mean optical properties simulating weakly to strongly absorbing/scattering tissue. (**c**) Simulated multispectral optoacoustic image (one wavelength presented). A circular grid is employed in the upper left part of the image for further analysis using eMSOT. (**d**) Original (green) and noisy (blue) simulated spectral response stemming from one pixel of (**c**). (**e-g**) Maps of *Eigenfluence* parameters $m_1$, $m_2$ and $m_3$, respectively obtained after inversion and interpolation. (**h-i**) $sO_2$ estimation using linear unmixing (**h**) and eMSOT (**i**). (**j**) Actual simulated $sO_2$ map. (**k**) $sO_2$ estimation error corresponding to all pixels of the analyzed area using conventional linear unmixing (red points) and the eMSOT method (blue points), sorted per depth. (**l**) Mean $sO_2$ error of linear unmixing (red) and eMSOT inversion (blue) corresponding to each simulated data-set tested. (**m**) Histogram of the mean $sO_2$ estimation error corresponding to eMSOT (blue) and linear unmixing (red) for all simulated data-sets tested. (**n**) Histogram of the relative $sO_2$ estimation error of linear unmixing as compared to eMSOT for all simulated data-sets tested and simulated tissue depths>5mm.

|  | Physiological range (30%-80% mean sO$_2$) | | | | 0%-30% mean sO$_2$ | 80%-100% mean sO$_2$ | Vessel network (30%-80% sO$_2$) | |
|---|---|---|---|---|---|---|---|---|
| μ$_a$$^{mean}$ (cm$^{-1}$) | [0.07-0.15] | | [0.2-0.3] | | [0.07-0.3] | [0.07-0.3] | [0.1, 0.2, 0.3] | |
| μ$_s$ $^{mean}$(cm$^{-1}$) | [7-11] | | [7-11] | | [7-11] | [7-11] | [7, 9, 11] | |
| Noise lvl. | 2.5% | 4.5% | 2.5% | 4.5% | 2.5% | 2.5% | 2.5% | 2.5% |
| Scale |  |  |  |  |  |  | 1-3 | 3-6 |
| Mean sO$_2$ error | 2.36% (4.54%) | 2.67% (4.65%) | 2.82% (7.9%) | 3.38% (7.9%) | 5.1% (15.6%) | 1.85% (11%) | 2.45% (5.83%) | 2.0% (4.4%) |
| % of pixels <10% error | 98.6% (89.4%) | 98.1% (89.1%) | 97.1% (70.8%) | 95.0% (70.4%) | 85.8% (38%) | 99.5% (56%) | 98.3% (81.7%) | 99.1% (87.5%) |
| % of pixels <15% error | 99.8% (97.2%) | 99.7% (97%) | 99.3% (85%) | 98.7% (84.8%) | 97% (57%) | 99.9% (74.9%) | 99.8% (93%) | 99.8% (96%) |

**Supplementary Table 1. Statistics of the eMSOT performance as evaluated on large simulated data-set (red corresponds to conventional linear unmixing).**

| Grid points | 12 | 30 | 56 | 108 |
|---|---|---|---|---|
| Av. computational speed (sec) | 1.8 sec | 10 sec | 52 sec | 487 sec |
| Mean sO$_2$ error | 3.16% | 2.74% | 2.5% | 2.36% |
| % of pixels <10% error | 95.9% | 97.7% | 98.1% | 98.5% |

**Supplementary Table 2. Statistics of the eMSOT performance as a function of grid density.** Statistics correspond to 108 simulated data-sets of μ$_a$$^{mean}$ ∈ [0.1 0.3] cm$^{-1}$, μ$_s$ $^{mean}$=10cm$^{-1}$ and mean sO$_2$ varying between 30%-80%.

**Supplementary Note 4: Validation of eMSOT with tissue mimicking blood phantoms**

Blood phantoms with controlled oxygenation levels were created for validating the eMSOT accuracy under experimental conditions where gold standard is available. Different blood sO$_2$ levels were created by adding different amounts of Sodium Dithionite (Na$_2$O$_4$S$_2$)[4], a chemical that allows for efficient deoxygenation of blood. Control experiments indicated that blood solutions in NaCl and intralipid could be stably retained at 100% sO$_2$ under no Na$_2$O$_4$S$_2$ addition and at 0% under 100 mg/g Na$_2$O$_4$S$_2$ addition. When Na$_2$O$_4$S$_2$ was added at a concentration of 2-4 mg/g, blood solutions were initially deoxygenated but would gradually change to higher oxygenation levels.

Different types of cylindrical (diameter 2cm) tissue mimicking solid blood phantoms were created consisting of 3%-5% blood in a solution of NaCl, intralipid (2%) and low temperature melting Agarose. Four different states of background blood oxygenation were formed though the administration of 100 mg/g Na$_2$O$_4$S$_2$ (corresponding to 0% sO$_2$ background), 3 mg/g Na$_2$O$_4$S$_2$, 4 mg/g Na$_2$O$_4$S$_2$ (corresponding to an unknown and spatially varying sO$_2$ in background) and 0 mg/g Na$_2$O$_4$S$_2$ (corresponding to 100% sO$_2$ background). A 3mm diameter insertion containing a sealed capillary tube filled with 20% blood at 0% sO$_2$ and 100% sO$_2$ was introduced at a depth of 5-8mm in each solid blood phantom. The phantoms were imaged using MSOT and the images were analyzed using the eMSOT method and conventional linear unmixing.

**Supplementary Fig. 5a-b** present the application of the eMSOT method in the case of a uniform phantom of 0% sO$_2$ and a phantom of 100% sO$_2$, respectively. **Supplementary Fig. 5c-d** present the sO$_2$ estimation error of the eMSOT method (blue dots) and linear unmixing (red dots) for all analyzed pixels sorted per imaging depth.

**Supplementary Fig. 5e-f** present the application of the eMSOT method in the case of an unknown, non-uniform sO$_2$ background phantom with an insertion of 0% sO$_2$ blood. The eMSOT grid is placed appropriately to cover the insertion area. **Supplementary Fig. 5g-h** present the initial spectrum in the insertion area (P(**r**,λ)) and the sO$_2$ estimation using linear unmixing (**g**) as well as the corrected spectrum (P$^{eMSOT}$(**r**,λ)) and sO$_2$ estimation using eMSOT method (**h**). **Supplementary Fig. 5i** summarizes the sO$_2$ estimation error of linear unmixing (red) and eMSOT method (blue) corresponding to the insertion area in the case of 8 different blood phantoms (4 different backgrounds and 2 different insertions per background). eMSOT offers higher accuracy with an sO$_2$ estimation that is typically less than 10%, as opposed to linear unmixing that can be associated with errors as high as 30%. Finally, **Supplementary Fig. 5i** presents the fitting residual of linear (red) and eMSOT unmixing (blue) in each case.

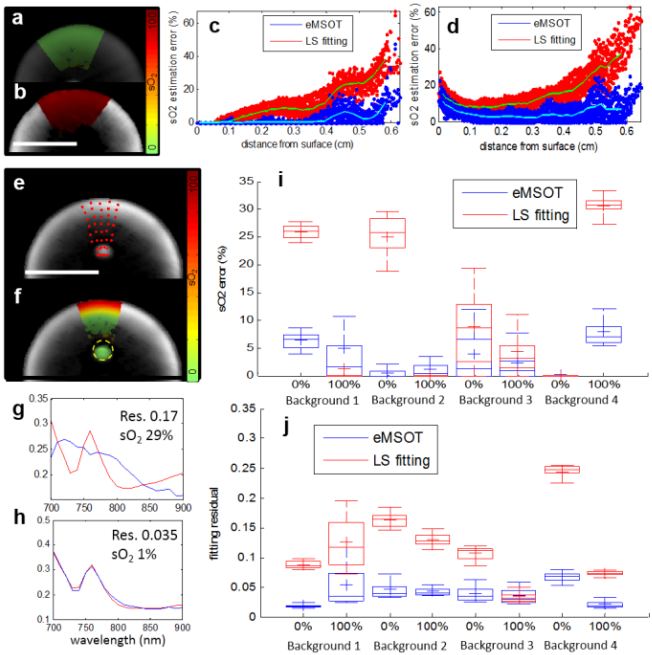

**Supplementary Figure 5. Validation of eMSOT using blood phantoms.** (**a, b**) eMSOT $sO_2$ estimation in the case of a uniformly deoxygenated blood phantom (**a**) and a uniformly oxygenated phantom (**b**). Scale bar, 1 cm. (**c, d**) $sO_2$ estimation error of eMSOT (blue dots) and linear unmxing (red dots) sorted per depth in the case of the deoxygenated phantom (**c**) and oxygenated phantom (**d**). (**e, f**) eMSOT grid application (**e**) and $sO_2$ estimation (**f**) in the case of a blood phantom containing an insertion of 0% $sO_2$. Scale bar, 1 cm. (**g, h**) Spectral fitting and $sO_2$ estimation in the insertion area using linear unmixing (**g**) and eMSOT (**h**). (**i**) Statistics on the $sO_2$ estimation error of eMSOT (blue) and linear unmixing (red) corresponding to the insertion region of eight different phantoms of varying background and target oxygenations. (**j**) Statistics on the fitting residual of eMSOT (blue) and linear unmixing (red) corresponding to the insertion region of eight different phantoms of varying background and target oxygenations.

**Supplementary Note 5: Application of eMSOT on experimental tissue images**

In experimental tissue data (muscle and tumor analysis) the prior ḿ$_1$ and ḿ$_3$ maps were computed as described in **Supplementary Note 2** by using a 3D light propagation model and 20 $sO_2$ maps corresponding to 20 consecutive MSOT slices (with a step size 0.5 mm) surrounding the central slice to be analyzed (**Supplementary Figure 6a**). This was performed in order to provide robust *Eigenfluence* prior estimates even in cases of substantial $sO_2$ variations in the 3D illuminated volume (MSOT illumination width ~ 1 cm). **Supplementary Figure 6b** presents the prior ḿ$_1$ map corresponding to an animal imaged *post-mortem* after $CO_2$ breathing.

eMSOT accuracy depends on the quality of the measured optoacoustic spectra in the grid area. For ensuring successful application, an image area of high intensity (high SNR) and fidelity (visually presenting no reconstruction artefacts e.g. due to ill ultrasound coupling) and typically corresponding to the central-upper part of the image (corresponding to the focal area of the ultrasound sensors and eliminating the possibility of reconstruction artefacts due to the limited angle of coverage) was selected for applying the eMSOT method. Upon manual segmentation of an area, a circular grid is automatically applied in the image domain (**Supplementary Figure 6c**). The grid point location is automatically updated so that the points occupy the highest intensity pixels in their local vicinity. Grid points that correspond to image values under a predefined threshold (i.e. red points in **Supplementary Figure 6c**) are excluded from the inversion. The measured optoacoustic spectra corresponding to the grid points are in the following used in the context of the constrained inversion algorithm described in **Methods** and **Supplementary Note 2** to obtain estimates of $m_1(\mathbf{r})$, $m_2(\mathbf{r})$ and $m_3(\mathbf{r})$ for each grid point $\mathbf{r}$. **Supplementary Figure 2d** presents the prior ḿ$_1(\mathbf{r})$ (blue line), the limits of search space (blue vertical lines) and the $m_1(\mathbf{r})$ estimated by the constrained inversion (red line) for a radial line of the grid in **Supplementary Figure 6c**.

Upon the estimation of $m_1(\mathbf{r})$, $m_2(\mathbf{r})$ and $m_3(\mathbf{r})$ in all grid points, the *Eigenfluence* maps for the intermediate grid points are computed by means of cubic interpolation (see **Methods**). **Supplementary Figure 6e, f** presents the $m_2$ (**e**) and $m_1$ (**f**) *Eigenfluence* maps corresponding to the same tissue area imaged under different physiological conditions, namely *post-mortem* after $CO_2$ breathing (left), *in-vivo* under 20%$O_2$ breathing (middle) and *in-vivo* under 100%$O_2$ breathing (right). While the $m_2$ map that corresponds mainly to tissue depth remains relatively unchanged under all three physiological conditions, $m_1$ that corresponds more to background tissue oxygenation presents substantial differences between the three different states. The *Eigenfluence* maps are used to correct for the wavelength dependence of light fluence in the selected tissue area (**Methods**) and in the following blood oxygen saturation maps are computed using non-negative constrained least squares fitting of the corrected eMSOT image with the spectra of oxy- and deoxy-hemoglobin (**Supplementary Figure 6g**). Pixels that are associated with a fitting residual above a certain threshold are excluded from the $sO_2$ maps.

After eMSOT inversion, the raw optoacoustic spectra (blue lines in **Supplementary Figure 6h** left) are decomposed into the element-wise product of the corrected normalized absorption spectra (blue lines in **Supplementary Figure 6h** middle) and the estimated light fluence spectra (**Supplementary Figure 6h** right). While linear fitting with the spectra of oxy- and deoxy-hemoglobin results in a high fitting residual and an inaccurate $sO_2$ estimation when applied on the raw optoacoustic spectra (red lines in **Supplementary Figure 6h** left), it results in a low fitting residual after eMSOT correction (red lines **Supplementary Figure 6h** middle) independently of tissue depth.

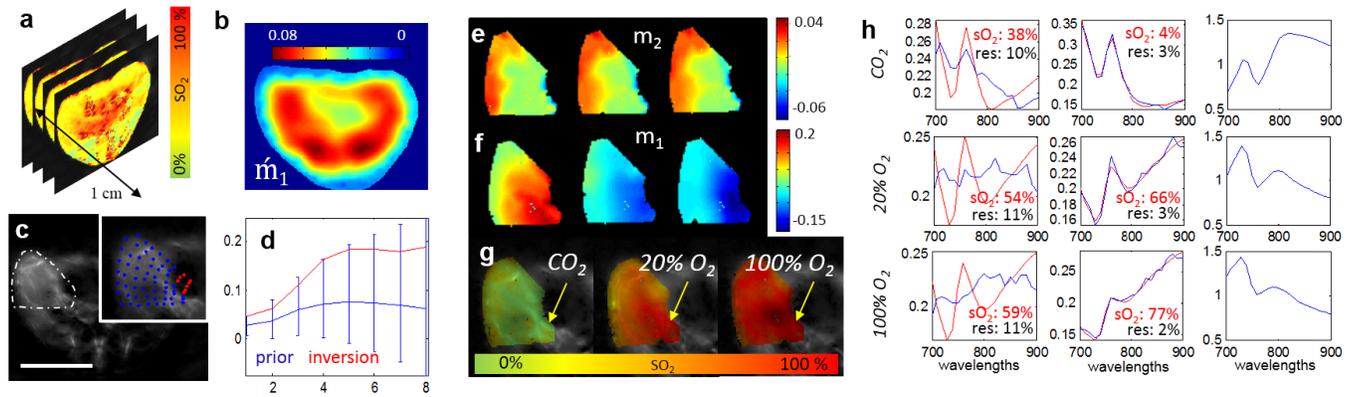

**Supplementary Figure 6. eMSOT application in experimental tissue images.** (**a**) Initial sO$_2$ maps corresponding to multiple MSOT slices surrounding the central slice to be analyzed. (**b**) Prior ṁ$_1$ map computed using a 3D light propagation model and the initial sO$_2$ maps as described in **Supplementary Note 2**. (**c**) Selection of a high intensity area in a well-reconstructed part of the image for the automatic application of a grid for eMSOT application. Scale bar, 1 cm. (**d**) Prior ṁ$_1$ (blue line), limits of search space (blue vertical lines) and estimated m$_1$ after eMSOT inversion, corresponding to a radial line of the grid in (**c**). (**e-g**) m$_2$ (**e**), m$_1$ (**f**) and sO$_2$ maps (**g**) computed after eMSOT inversion for the same tissue area under three different physiological conditions. (**h**) Original optoacoustic spectra (P(**r**,λ); left, blue), eMSOT spectra (P$^{eMSOT}$(**r**,λ); middle, blue) and estimated spectrum of light fluence (right) corresponding to a deep tissue point (yellow arrow in **g**). The spectral fitting with the spectra of oxy- and deoxygenated haemoglobin (red) and the estimated sO$_2$ value and fitting residual are also presented in each case.

## Supplementary Note 6: Imaging tumor hypoxia with eMSOT and histological validation

N=8 mice, bearing orthotopically implanted 4T1 mammary tumors were imaged with MSOT at transverse slices in the lower abdominal area (schematic representation in **Supplementary Figure 7a**). **Supplementary Figure 7b** presents an anatomical optoacoustic image showing a slice which corresponds approximately to the central section of the tumor. The tumor region (upper right part of the image) can be recognized as it displays an enhanced contrast and different anatomic characteristics as compared to the symmetric normal tissue region. The tumor region is manually segmented (dashed segmentation line, **Supplementary Figure 7b**). The eMSOT grid is set to cover the tumor area as well as adjacent normal tissue (**Supplementary Figure 7b right**).

After MSOT imaging, the mice were sacrificed and prepared for histological analysis. A subset of the mice (n=4) were examined for tumor and tissue anatomy. Following MSOT acquisition, the mice were frozen and the lower abdominal region containing the tumor mass (dashed lines in **Supplementary Figure 7c**) was cryosliced in transversal orientation, similar to the one of MSOT imaging (see **Supplementary Figure 7a**). True color images of the whole body, including the tumor mass, were obtained and histological slices derived thereof were isolated for H&E staining. **Supplementary Fig. 7d-g** presents an anatomical optoacoustic image at the central tumor cross-section (**d**), the corresponding cryoslice true color photography (**e**), H&E tumor staining (**f**) and eMSOT sO$_2$ analysis (**g**). The cryoslice true color photography displays the tumor heterogeneity, presenting sub-regions with prominent red color (marked in **Supplementary Fig. 7e** with an asterisk). These central necrotic areas, appearing to be suffused with blood, spatially correlate to the central hypoxic region in the core of the tumor as identified in the eMSOT image (**Supplementary Figure 7g;** marked with an asterisk). Central necrotic areas could be confirmed by H&E staining (**Supplementary Figure 7f**).

Another subset of the mice (n=4) was examined for functional characterization of the tumors through CD31/Hoeachst33342/Pimodinazole histological staining. Throughout this process, the tumors were excised and the 3D orientation of the tumor with regard to the MSOT image was retained (**Supplementary Fig. 7h, lower picture**). In the following, the excised tumors were sectioned and ~8 μm thick slices were immunohistochemically stained for studying micro-vascularization (CD31 staining) and cellular hypoxia (Pimonidazole staining). Vascular perfusion was determined following Hoechst33342 detection.

**Supplementary Fig. 7i** presents the eMSOT sO$_2$ estimation of two tumors presenting different levels of oxygenation. The tumor areas, as identified by the anatomical images, are segmented with a yellow dashed line. The average sO$_2$ levels of the central tumor areas (blue dashed rectangle) are further displayed in the image. The corresponding CD31 staining, as shown in **Supplementary Fig. 7j** reveals a dense tumor microvasculature in both tumors. This might explain the high tumor contrast in optoacoustic imaging. Hoechst 33342 staining (**Supplementary Fig. 7k**) reveals substantial differences in the perfusion patterns of the two tumors, with the first tumor appearing to be perfused both in the boundary (grey dashed box) and the core (green dashed box). In an effort to quantify the perfusion patterns, the ratio of the Hoechst image intensity in the core vs the boundary was computed (intensity ratio 48%). The second tumor displays less perfusion in the core, as compared to the boundary (intensity ratio 19%). This finding indicates less functionality of the microvasculature in the core, which might explain the lower eMSOT sO$_2$ values as compared to the first tumor. The less perfused tumor areas (dark areas in **k**) appear spatial congruence with the areas of reduced blood oxygenation revealed by eMSOT (**i**). The non-perfused tumor areas further appear spatially correlated to cell hypoxia as identified by Pimonidazole staining (**l**, green). Cell hypoxia, as determined by Pimonidazole staining, may be a consequence of both, perfusion hypoxia (revealed by Hoechst33342 and eMSOT) and also diffusion hypoxia, which does not display eMSOT signal. Although, due to technical reasons, it may be challenging to achieve exact co-registration between *in-vivo* eMSOT tumor images and *ex-vivo* histology, the presented histological analyses demonstrate the ability of eMSOT to detect perfusion related hypoxia within solid tumors.

Furthermore, clear discrimination of different levels of hypoxia within single tumors, as well as intratumoral hypoxia-related heterogeneity could be demonstrated.

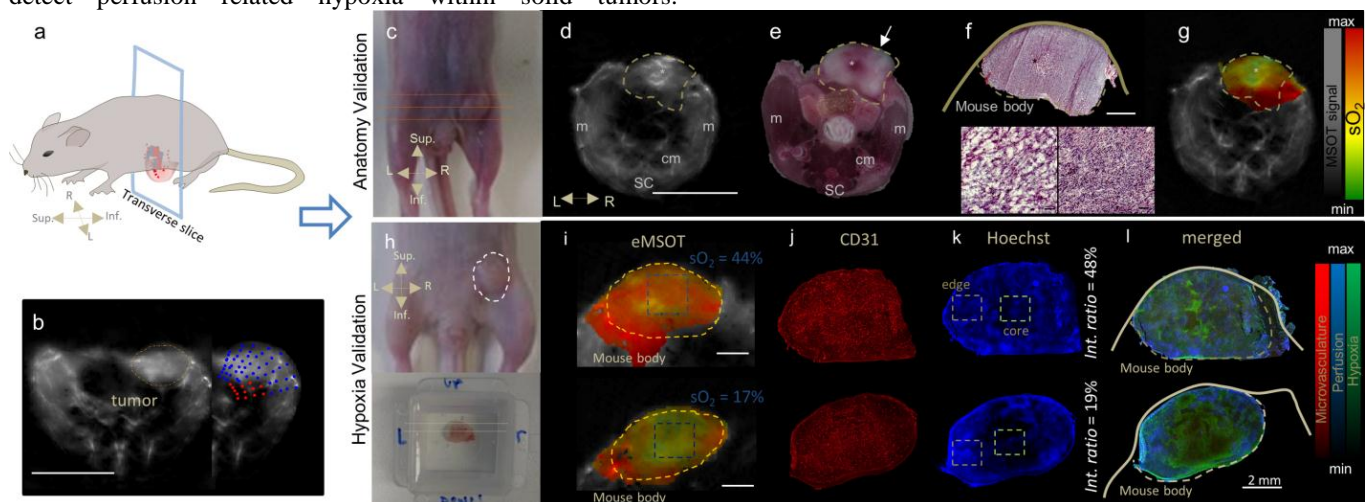

**Supplementary Figure 7. Histological validation of tumor imaging and co-registration.** (**a**) Schematic representation of MSOT imaging at a transversal slice within the tumor area (**b**) Cross-sectional optoacoustic image at a central tumor transversal slice. The tumor region is segmented with a dashed line. The eMSOT grid is further presented (blue and red dots). (**c**) Image of the lower abdominal area displaying the orthotopic mammary tumor. Dashed lines present the orientation of cryoslicing and MSOT imaging. (**d-g**) Anatomical optoacoustic image (**d**; Scale bar, 1cm) and the corresponding cryosliced color photography (**e**), H&E staining of the tumor region (**f**; Scale bar, 2mm) and eMSOT $sO_2$ analysis of the tumor area (**g**). (**h**, lower) Excised tumor used for functional staining. Yellow dashed lines indicate the slicing orientation. (**i-l**) Examples of a highly perfused (upper row) and low perfused (lower row) tumor analysed with eMSOT for $sO_2$ estimation (**i**), CD31 staining (**j**), Hoeachst33342 staining (**k**), and merged with Pimonidazole staining (**l**). Scale bar, 2mm. The tumor margins are presented in (**i**) indicated by yellow dashed lines. Blue dashed rectangles indicate a region in the tumor core, the average $sO_2$ values of which are displayed on the upper right. The intensity ratio of Hoechst33342 staining was calculated by dividing the mean intensity value in the tumor core (green dashed rectangle in (**k**)) over the one in the tumor boundary (grey rectangle in (**k**)).

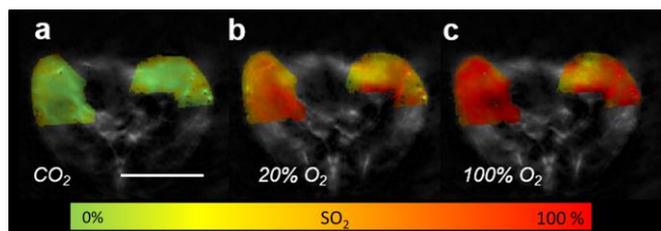

**Supplementary Figure 8. Comparison of healthy tissue and tumor $sO_2$ measurements under a breathing challenge.** (**a-c**) Healthy tissue (left) and tumor (right) $sO_2$ estimation post-mortem after $CO_2$ breathing (**a**) and *in-vivo* under 20%$O_2$ (**b**) and 100% $O_2$ breathing (**c**).